\documentclass[a4paper,12pt]{article}

\usepackage{graphicx}
\usepackage{amssymb}
\usepackage{amsmath}
\usepackage{color}
\usepackage[labelfont=bf]{caption} 

\usepackage{palatino}
\usepackage[T1]{fontenc}

\definecolor{violet}{rgb}{0.7,0,0.7}

\def\url#1{\textcolor{blue}{\underline{#1}}}	

\definecolor{oneblue}{rgb}{0,0,0.65}
\def\b#1{\textcolor{oneblue}{#1}} 

\newcommand{\degC}{${\,}^\circ C{\,}$}

\begin{document}


\title{Intracellular impedance measurements reveal non-ohmic 
properties of the extracellular medium around neurons}

\author{Jean-Marie Gomes$^{*1}$,Claude B\'edard$^{*1}$,
  Silvana Valtcheva$^2$, \\
  Matthew Nelson$^3,4$, Vitalia Khokhlova$^1$, Pierre Pouget$^3$, \\
  Laurent Venance$^{2 \ddag}$, Thierry Bal$^{1 \ddag}$ and
  Alain Destexhe$^{{1\ddag \dag}}$ \\ \ \\
  \small 1: Unit\'e de Neurosciences, Information et Complexit\'e (UNIC), \\
  \small   Centre National de la Recherche Scientifique (CNRS), \\
  \small   Gif-sur-Yvette, France; \\
  \small 2: Centre Interdisciplinaire de Recherche en Biologie (CIRB)  \\  \small CNRS UMR 7241 - Inserm U1050, Coll\`ege de France, Paris, France;\\
  \small 3: Institut du Cerveau et de la Moelle Epini\`ere (ICM), \\
  \small CNRS UMR 7225 - Inserm UMRS 975, Universit\'e Pierre et Marie Curie,\\
  \small   H\^opital de la Salp\'etri\`ere, Paris, France; \\
  \small 4: Present address: Cognitive Neuroimaging Unit, INSERM U992, \\
  \small   NeuroSpin Center, Gif-sur-Yvette, France \\
  \small *~: co-first authors;  \\
  \small \ddag: These authors jointly co-supervised this work \\
  \small \dag: Corresponding author, email: destexhe@unic.cnrs-gif.fr \\
  \small \ \\
    {\it Biophysical Journal}, in press.}
\maketitle

\vspace{5mm}

\noindent {\bf Running title:} Impedance measurements in cerebral cortex

\vspace{5mm}

\noindent {\bf Corresponding author:} Alain Destexhe, UNIC, CNRS, 1 Avenue
de la Terrasse, \\ 91190 Gif sur Yvette, France. \ Tel: +33 1 69 82 34 35 \
destexhe@unic.cnrs-gif.fr

\vspace{5mm}

local field potential, extracellular

\vspace{5mm}

\noindent {\bf Conflict of interest:} None

\clearpage


\section*{Abstract}

The electrical properties of extracellular space around neurons are
important to understand the genesis of extracellular potentials, as
well as for localizing neuronal activity from extracellular
recordings.  However, the exact nature of these extracellular
properties is still uncertain.  We introduce a method to measure the
impedance of the tissue, and which preserves the intact cell-medium
interface, using whole-cell patch-clamp recordings {\it in vivo} and
{\it in vitro}.  We find that neural tissue has marked non-ohmic and
frequency-filtering properties, which are not consistent with a
resistive (ohmic) medium, as often assumed.  
The amplitude and phase profiles
of the measured impedance are consistent with the contribution of
ionic diffusion.  We also show that the impact of such
frequency-filtering properties is possibly important on the genesis of
local field potentials, as well as on the cable properties of neurons.
The present results show non-ohmic properties of the extracellular
medium around neurons, and suggest that source estimation methods, as
well as the cable properties of neurons, which all assume ohmic
extracellular medium, may need to be re-evaluated.


\section*{Introduction}

The genesis and propagation of electric signals in brain tissue depend
on its electric properties, which can be simply resistive (ohmic) or
more complex, such as capacitive, polarizable or diffusive. The exact
nature of these electric properties is important, because
non-resistive media will necessarily impose frequency-filtering
properties to electric signals~\cite{Herreras,Buzsaki}, and therefore
will influence any source localization.  These electric properties
were measured using metal electrodes, {which provided
  measurements suggesting} that the brain tissue is essentially
resistive~\cite{Ranck,Nicho2005,Logothetis}.  However, {the
  electrical behavior of tissue and electrodes can be easily
  confused~\cite{Schwan1968}; efforts in the direction of
  distinguishing or separating them abound
  (\cite{Geddes,McAdams,Schwan1966}).}  Another experimental approach
{using very low-impedance probes} revealed marked frequency
dependence of conductivity and permittivity~\cite{Gabriel,Wagner}.
Indirect evidence for non-resistive media were also
obtained~\cite{BedDes2010,Deg2010,Nelson2013}, and also indicated a
marked frequency dependence.

To explain these discrepancies, we hypothesize that the apparently
contradictory results are due to the fact that different measurement
methods were used.  The use of metal electrodes represents a
non-physiological interface for interacting with the surrounding
tissue, while in reality, neurons interact with the extracellular
medium by exchanging ions through membrane ion channels and pumps.  To
respect as much as possible these natural conditions, we have measured
the tissue impedance using a neuron as a current source, thereby
respecting the natural interface. This intracellular measurement
provides a measurement of a global cell impedance, which contains the
membrane impedance and the impedance of the extracellular medium.
This global intracellularly-measured impedance can also be defined as
the impedance as seen by the cell.  

{Figure~\ref{principle} illustrates this concept and the
  recording setup needed to record this global impedance
  intracellularly, {\it in vitro} (Fig.~\ref{principle}A) or {\it in
    vivo} (Fig.~\ref{principle}B).  Fig.~\ref{principle}C gives two
  circuit configurations for this system, emphasizing three
  impedances: \b{$Z_i$} is the impedance of the intracellular medium
  (cytoplasm), \b{$Z_e$} is the extracellular impedance, and
  \b{$Z_{RC}$} is the membrane impedance, represented by a simple
  resistance-capacitance (RC) circuit (left), or a more complex
  circuit including dendritic compartments all described by different
  RC circuits (right).  An intracellular electrode will measure a
  global combination of these impedances, In the following of the
  text, we will call this global impedance the ``global intracellular
  impedance''.}

{A central point of our study is that this global intracellular
  impedance is different} than the electrical impedances measured by
metal electrodes, which we refer here as ``metal-electrode
impedance''.  We will investigate whether the
intracellularly-measured impedance reveals more complex electrical
properties than with metal electrodes, which could possibly explain
the discrepancies {described above}.  The global intracellular
impedance provides not only a realistic estimate of the electrical
properties the extracellular medium, but it is also closer to the
natural conditions, and could be a useful physical parameter to
determine {a more precise} source localization of cerebral
electric signals, and to model the propagation of electrical signals
in the extracellular space or in dendritic trees, as we illustrate
here.


\section*{Materials and Methods}

\subsection*{Animals}

Maintenance, surgery and all experiments were performed in accordance
with the local animal welfare committees (Center for Interdisciplinary
Research in Biology and EU guidelines, directive 2010/63/EU, and
regional ethics committee "Ile-de-France Sud'' (Certificate 05-003)).
Every precaution was taken to minimize stress and the number of
animals used in each series of experiments. Animals were housed in
standard 12 hours light/dark cycles and food and water were available
ad libitum.

\subsection*{\emph{In vitro} electrophysiology}

\textbf{Brain slice preparation.}\\
{\textit{Visual cortex.} 300 $\mu$m-thick coronal brain slices of
  juvenile mice ($P_{12-16}$ Swiss mice bred in the CNRS Animal
  Care facility, Gif-sur-Yvette ; French Agriculture Ministry
  Authorization: B91-272-105)} were obtained with a Leica VT 1200S
microtome (Leica Biosystems, Wetzlar, Germany). Slices were prepared
at 4\degC in the following medium (in mM): choline chloride 110, KCl
2.55, NaH$_2$PO$_4$ 1.65, NaHCO$_3$ 25, dextrose 20, CaCl$_2$ 0.5,
MgCl$_2$ 7. Slicing started 2 mm from the posterior limit of olfactory
bulb, and ended 3.9mm further.  Before recordings, slices were
incubated at 34\degC in artificial cerebro-spinal fluid (ACSF) of the
following composition (in mM): NaCl 126, KCl 3, NaHC$O_3$ 26,
NaH$_2$P$O_4$ 1.25, myo-inositol 3, sodium pyruvate 2, L-ascorbate de
sodium 0.4 ; dextrose, 10.  The slicing and recording solution was
bubbled with 95\% O$_2$ and 5\% CO$_2$, for a final pH of 7.4.

\noindent{\textit{Dorsal striatum.}  Horizontal brain slices with a
  thickness of 330 $\mu$m were prepared from rats ($P_{23-30}$ OFA
  rats (Charles River, L'Arbresle, France), using a vibrating blade
  microtome (VT1200S, Leica Microsystems, Nussloch, Germany). Brains
  were sliced in a 95\% O$_2$ / 5\% CO$_2$-bubbled, ice-cold cutting
  solution containing (in mM): NaCl 125, KCl 2.5, glucose 25,
  NaHCO$_3$ 25, NaH$_2$PO$_4$ 1.25, CaCl$_2$, MgCl$_2$ 1, pyruvic acid
  1, and then transferred into the same solution at 34\degC.}

\noindent\textbf{Electrophysiological recordings.}\\
Patch-clamp recordings in pyramidal cells of visual cortex from mice
were performed as followed.  {Slices were superfused at 2 mL/min
  with the same ACSF solution that was used for incubation. Bath
  temperature of the submerged chamber was maintained at 34 \degC
  using a TC-344B temperature controller (Warner Instruments Company,
  Hamden, CT, USA).}  Neurons in slices of the {mouse visual
  cortex ($P_{12-16}$)} were identified with an upright microscope
(Axioscope FS, Zeiss, Germany), an infrared camera (C750011, Hamamatsu
Photonics, Japan) and an infrared filter.  Patch-clamp in the
whole-cell current clamp configuration in layer V pyramidal neurons
was performed simultaneously with an extracellular recording using a
{3 M$\Omega$ patch pipette} (Fig.~\ref{principle}A). The latter
was located within a close vicinity (\b{$\approx 30~\mu m$}) of the
patched cell. All results were indifferent to the distance between the
reference electrode and the patched cells, as potential variations on
the extracellular electrode did not excede 1\% of the variations of
the intracellular potential.  Borosilicate pipettes (1B150F4, World
Precision Instruments, Inc., Sarasota, Florida, U.S.) of 5-8 M$\Omega$
impedance were used for whole-cell recordings and contained the
intracellular solution (in mM): Hepes 10, EGTA 1, K-gluconate 135,
MgCl$_2$ 5, CaCl$_2$ 0.1, with osmolarity of 308 mOsm and a pH of 7.3.
Serial resistance was not compensated for.  Recordings were performed
with a Multiclamp 700B amplifier (Axon Instruments Inc., California,
U.S.), filtered at 10 kHz with a built-in Bessel filter, and sampled
at 25 kHz. Data acquisition and stimulation were performed with a
National Instruments BNC 2090 A card, and the software ELPHY (G.
Sadoc, CNRS, UNIC, France).

{Patch-clamp recordings in medium-sized spiny neurons of dorsal
  striatum from rats were performed as previously
  described~\cite{Paille2013}. Briefly, borosilicate glass pipettes of
  6-8M impedance contained for whole-cell recordings (in mM):
  K-gluconate 105, KCl 30, HEPES 10, phosphocreatine 10, ATP-Mg 4,
  GTP-Na 0.3, EGTA 0.3 (adjusted to pH 7.35 with KOH). The composition
  of the extracellular solution was (mM): NaCl 125, KCl 2.5, glucose
  25, NaHCO$_3$ 25, NaH$_3$PO4 1.25, CaCl$_2$ 2, MgCl$_2$ 1, pyruvic
  acid 10, bubbled with 95\% O$_2$ and 5\% CO$_2$. Signals were
  amplified using EPC10-3 amplifiers (HEKA Elektronik, Lambrecht,
  Germany). All recordings were performed at 34\degC using a
  temperature control system (Bath-controller V, Luigs \& Neumann,
  Ratingen, Germany) and slices were continuously superfused at 2
  ml/min with the extracellular solution. Slices were visualized on an
  Olympus BX51WI microscope (Olympus, Rungis, France) using a 4x/0.13
  objective for the placement of the stimulating electrode and a
  40x/0.80 water-immersion objective for localizing cells for
  whole-cell recordings. During the experiment, individual striatal
  and cortical neurons were identified using infrared-differential
  interference contrast microscopy with a CCD camera (Optronis VX45;
  Kehl, Germany). Serial resistance was not compensated for.
  Current-clamp recordings were filtered at 2.9 kHz and sampled at
  16.7 kHz 
  using the Patchmaster v2x73 program (HEKA
  Elektronik). Stimuli in current-clamp mode underwent a high cut
  10kHz filter before being applied. Recordings were made with EPC
  10-3 amplifiers (HEKA Elektronik; Lambrecht, Germany) with a very
  high input impedance ($1~T\Omega$) to ensure there was no
  appreciable signal distortion imposed by the high impedance
  electrode (Nelson et al., 2008). Sinusoidal stimuli were then
  introduced in whole-cell patch-clamp to the patched cell.}

This configuration enables estimating the extracellular impedance,
according to the circuit displayed in Fig.~\ref{principle}C. To this
end, a white noise current stimulus was injected into the recorded
cell and the impedance was calculated based on this current injection
{(see Fig.~\ref{invitro}A).}  {20 to 120 seconds of Gaussian
  white noise with zero mean and 100 pA variance was injected (results
  were similar for 30 and 200 pA variance). For each cell, we injected
  15 to 30 times the same sequence of white noise (``frozen noise'')
  and averaged the measured voltages. This enhanced the signal to
  noise ratio without altering the results. Figure~\ref{invitro}A
  (right panel) shows the very low autocorrelation of the injected
  currents, being thus a good approximation of white noise.  To verify
  that the same results can be obtained using a different protocol, we
  also performed slice experiments using sinusoid current stimuli, at
  different frequencies. This set of experiments was performed
  following the methods previously
  published~\cite{Nelson2013,Nelson2008}.  Namely, sine waves of 12
  different frequencies were tested, varying approximately evenly on a
  logarithmic scale ranging from 6 Hz to 926 Hz. Up to 500 traces of
  100 to 1500 ms in length were averaged before recording the data to
  disk for offline analyses. Longer stimulus lengths and more traces
  were necessary for the low frequency stimuli.  The order of the
  presentation of the frequencies was randomized.  Stimuli were
  introduced with the patch electrode in {current-clamp} mode.
  The injected current amplitudes ranged from 200 to 300 pA.  Before
  conducting experiments, we verified via control recordings with an
  external signal generator in the bath without a slice that any
  amplitude changes or phase shifts in the recordings across
  frequencies induced by the amplifier and recording hardware were
  negligible~\cite{Nelson2008}.}

\subsection*{\emph{In vivo} electrophysiology} 

{\textbf{Surgical preparation of animals.}}\\
Adult rats (P40-90) were placed in a stereotaxic apparatus
(Unimecanique, Asni\`eres, France) after anesthesia induction with a
400mg/kg intra-peritoneal injection of chloral hydrate (Sigma-Aldrich,
Saint-Quentin Fallavier, France). A deep anesthesia maintenance was
ensured by intra-peritoneal infusion on demand of chloral hydrate
delivered with a peristaltic pump set at 60mg/kg/hour turned on one
hour after induction.  Proper depth of anesthesia was assessed
regularly by testing the cardiac rhythm, EcoG activity, the lack of
response of mild hindpaw pinch and the lack of vibrissae movement. The
electrocardiogram was monitored throughout the experiment and body
temperature was maintained at 36.5\degC by a homeothermic blanket.

Two craniotomies were performed, one for the insertion of a reference
electrode in the somatosensory cerebral cortex (layer2/3) and one to
allow the whole-cell recordings in the contralateral cortex (layer 5).
For whole-cell recordings, a 2x2 mm craniotomy was made to expose the
left posteromedial barrel subfield at the following coordinates:
posterior 3.0-3.5 mm from the bregma, lateral 4.0-4.5 mm from the
midline. The dura was opened and the craniotomy was filled with low
melting point paraffine after each time lowering a recording pipette.
To increase recording stability the {cisterna magna} was drained.

\textbf{Electrophysiological recordings}

Borosilicate glass pipettes of \b{$5-8M\Omega$} impedance for blind
whole-cell recordings contained (in mM): K-gluconate 105, KCl 30,
HEPES 10, phosphocreatine 10, ATP-Mg 4, GTP-Na 0.3, EGTA 0.3 (adjusted
to pH 7.35 with KOH). Signals were amplified using an EPC10-plus-2
amplifier (HEKA Elektronik). Series resistance was not compensated
for. Current-clamp recordings were filtered at 2.5 kHz and sampled at
5 kHz 
using the Patchmaster v2x32 program (HEKA Elektronik).
Whole-cell recordings were performed in pyramidal cells of the
somatosensory cortex in layer IV/V (depth from the dura: 0.8-1.2 mm)
(Fig.~\ref{principle}B). Recorded cells were identified as pyramidal
cells {according to} their characteristic spiking pattern.  The
reference was a silver wire placed in the contralateral hemisphere.
Note that for both \textit{in vivo} and \textit{in vitro} experiments, the reference
electrode is passive, just measuring the extracellular voltage, and
thus the exact nature of this reference is not critical.
{Accordingly, the same configuration using a silver microwire
  gave similar results in vitro (not shown).}

It is important, however, that the reference electrode be placed in
the brain tissue, so that the estimated extracellular impedance is not
influenced by other tissues. Thus, as in the \textit{in vitro}
experiments, this intracellular-extracellular configuration enables
estimating the extracellular impedance. An important difference with
\textit{in vitro}, is that \textit{in vivo} the current can flow more
freely in 3 dimensions, and is closer to natural conditions. Another
difference is that \textit{in vivo}, the system is not silent but
displays prominent spontaneous activity. To limit this contribution,
we have used a ``frozen noise'' protocol, where identical sequences of
white noise stimuli were {injected repetitively, and the
  sequences averaged}.

{The main purpose of this experiment was to estimate
  \b{$Z_{eq}$}, the equivalent impedance between the Ag-AgCl electrode
  and the ground. In the case of a simple, single-compartment neuron,
  it can be formally defined as \b{$Z_{eq} = Z_i + Z_{RC} + Z_e$}.}
This was achieved by applying two protocols of subthreshold current
injection:
\begin{enumerate}
\item The ``frozen noise'' protocol consisting in the same template of
  20 seconds of a white noise current (repeated 50 times {with
    2~s intervals between stimulations and averaged}).  Sequences in
  which spikes were induced were discarded.
\item Sinusoidal current at fixed frequencies ranged from 6 Hz to 926
  Hz (similar to those used in {\it in vitro} experiments). The order of the
  presentation of the frequencies was randomized.
\end{enumerate}

For the frozen noise and sinusoidal stimuli, the injected currents
were tuned for each neurons to evoke voltage response of magnitude
ranged between 2 and 6 mV. Note that in some experiments we injected a
hyperpolarizing current ({of maximum amplitude $- 150$ pA}) to
prevent suprathreshold activity during application of stimuli.

\subsection*{Analysis}

All analyses were performed using Python (Python Software Foundation,
Wolfeboro Falls, New Hampshire, USA), the Scipy Stack and Spyder
(Pierre Raybaut, The Spyder Development Team).

\subsubsection*{\emph{In vitro} and \emph{in vivo} patch-clamp - sine
  waves experiments}

For each frequency and current intensity, the recorded voltages and
injected intensities were averaged and fit with sines using the
optimize package included in Scipy. The adequation between data and
the fitted sine waves was checked by human intervention for every set
of data. The voltage and current were represented respectively as
\b{$V(t,f) = V_0 sin(2 \pi f t + \phi_v)$} and \b{$I(t,f) = I_0 sin(2
  \pi f t + \phi_i)$}). The impedance for a given frequency was thus
given by \b{$Z(f) = \frac{V_0}{I_0} e^{i(\phi_v - \phi_i})$}.

\subsubsection*{\emph{In vitro} and \emph{in vivo} patch-clamp - white
  noise experiments}

Several models were fit to the experiments. First, we used a purely
resistive model (Fig. \ref{principle}, bottom left) in which the intracellular impedance \b{$Z_i$}
can be considered zero and the extracellular impedance \b{$Z_e$} was a small resistance ($R_e$).
The equivalent impedance is thus given by:
\begin{eqnarray}
\b{Z_{eq, 1}(\omega) =
  \frac{R_m}{1+i\omega \tau_m} + R_e}
	\label{eqresistive}
\end{eqnarray}
Second, we combined \b{$Z_i$} and \b{$Z_e$} as a diffusive term.
Rather than trying to find an elusive general solution for the usual
Nernst-Planck equation~\cite{Pods}, we used a {first-order}
approximation for ionic diffusion. The impedance of an electrolyte
showing non-negligible ionic diffusion was derived by
Warburg~\cite{Warburg1,Warburg2}, and yielded a modulus scaling in
\b{$1/\sqrt{\omega}$} and a constant phase. Similar derivations have
been performed in different
symmetries~\cite{Bisquert,BedDes2009,BedDes2011}. Note that the latter
derivations model the impedance of ionic accumulations close to the
membrane, by using a first-order approximation of the electric
potential generated by ionic species following Boltzmann
distributions.

This ``diffusion'' impedance has been observed
  experimentally (reviewed in ref.~\cite{Geddes}), and can be modeled
  in spherical symmetry by two components scaling the modulus and
  phase (\b{$A$} and \b{$B$}), and a cutoff frequency \b{$f_W =
    \omega_W/2\pi$}:
\begin{eqnarray}
	\b{Z_W(\omega) = \frac{A + iB}{1 + \sqrt{\omega/\omega_W}}}
	\label{Zwarburg}
\end{eqnarray}

This leads to the following expression for the equivalent impedance:
\begin{eqnarray}
\b{Z_{eq, 2}(\omega) = \frac{R_m}{1+i\omega \tau_m} +
  \frac{A + iB}{1 + \sqrt{i\frac{\omega}{\omega_W}}}}
\label{eqdiffusive}
\end{eqnarray}

To take into account the fact that some of the current can flow
through the dendrites of the cell, we define a dendritic input
impedance \b{$Z_d$} (see Fig.~\ref{principle}D, right), namely the
impedance of the dendritic tree seen by currents leaving the soma.
These currents will flow downwards the gradient potential from the
intracellular potential (\b{$V_{intra}$}) to the reference
(\b{$V_{ref}=0$}). Thus, \b{$Z_d$} is defined by
\b{$V_{intra}/i_d^g$}, where \b{$i_d^g$} is the generalized axial
current in the dendrite at the level of the soma, and \b{$V_{intra}$}
is the intracellular potential at the soma. Note that we consider here
\b{$V_{intra}$} as not necessarily equal to the {transmembrane
  potential ($V_m$)} because we take \b{$V_e$}, the potential of the
extracellular medium, into account.  We then consider separately the
resistance and capacitance of the soma (\b{$R_s$},\b{$C_s$}) and the
impedance of the dendrite. The equivalent impedance is then:
\b{$Z_{eq, 3-4}(\omega) = \frac{Z_d Z_s}{Z_d + Z_s} + Z_i + Z_e$},
where \b{$Z_s(\omega) = \frac{R_s}{1 + i\omega\tau_m}$} is the
impedance of the soma, {\b{$Z_{eq, 3}$} and \b{$Z_{eq, 4}$} are the impedances for
resistive and diffusive media respectively}. A description
of these different models can be found in the next section;
{parameters for each of these models are listed in
  Table~\ref{parameters}.}

\subsubsection*{Fitting models with and without dendrites}

Two types of models were fit to the experimental measurements, as
illustrated in Fig.~\ref{principle}C: a single-compartment model, and
a model including a dendritic segment.  Dendritic filtering has been
proposed to explain the frequency-dependence of
LFPs~\cite{Pettersen,Linden}; thus, current flowing in the dendrites
could be involved in shaping the measured impedance, which
would in turn influence the frequency-dependence of local field
potentials. We tested this possible influence by considering models
that include an equivalent dendritic compartment, which has been shown
to be electrically equivalent to a full dendritic
tree~\cite{Rall1962,Rall1995}, leading to a ``ball-and-stick'' type
model (see right circuit in Fig.~\ref{principle}C).

{In order to fit the models to experimental data, several
  traditional fitting methods were tested (e.g. Newton-Gauss,
  Levenberg-Marquardt, conjugate gradient, simplex...) but were
  plagued with three main problems: a long computation time (with 4
  parameters or more), a tendency to get trapped in local minima, and
  an extreme sensitivity to the first estimation of model parameters.
  We thus developed a probabilistic, non-iterative method that had
  none of these problems.}

{We proceeded as follows: 1. For each parameter, we defined a
  domain of acceptable values, keeping it very large (too much
  restriction on a parameter is a symptom of analytical bias). 2. We
  drew random sets of parameters {(between 500 and 5000)}, e.g
  \b{$P_j = (R_m, C_m, A, B, f_W...)$}, and computed the theoretical
  impedance spectra they predicted with a given model. 3. We computed
  the squared error \b{$E_j(P_j)$} between the theoretical and
  measured impedances. {The error was computed as the sum of
    squared errors on real and imaginary parts.}. 4.  The couple
  \b{$(P_j, E_j(P_j))$} that had the smallest errors over all tries
  was kept as the best fit.}

{Random drawing removed the sensitivity to local minima and
  initial parameters that is intrinsic to traditional iterative
  methods. It allowed a thorough exploration of the parameter space,
  and with high reliability {and acceptable efficiency.
    Empirically, this method was found to be much faster than a
    systematic exploration, but as reliable.}}

\subsubsection*{Modeling the contribution of dendrites}

To model the impedance of the cell including the contribution of
dendrites, we use the generalized cable formalism (FO model in
ref.~\cite{BedDes2013}), {which reads:
\b{
\begin{equation}
      \lambda^2\frac{\partial^2 V_m(x,\omega)}{\partial x^2} = \kappa^2 V_m(x,\omega)
\label{eq10}
\end{equation}}
where
\b{
\begin{equation}
\left \{
\begin{array}{ccc}
 \lambda^2 & = & \frac{r_m}{\bar{z}_i} \\
 \kappa^2  & = & 1+i\omega\tau_m
\end{array}
\right . ~ ,
\label{gen}
\end{equation}}
for a cylindric compartment.  Here, the quantity \b{$\bar{z_i}$} is an
equivalent impedance, which depends on the model considered.
\b{$\bar{z_i} = r_i + r_e$} for the standard (resistive) cable model,
defined from the intracellular and extracellular resistivities, \b{$r_i$} 
and \b{$r_e$}, respectively.  In the case of a frequency dependent 
impedance, \b{$\bar{z_i}$}  is more complex and is given by 
\b{$$\bar{z_i} = z_i / [1+\frac{z_e^{(m)}}{r_m}(1+ i\omega \tau_m)] ~ ,$$} 
where \b{$z_i$} and \b{$z_e$} are the intracellular (cytoplasm) and 
extracellular impedance densities, respectively, \b{$r_m$} is the 
membrane resistivity and \b{$\tau_m$} is the membrane time constant.
The estimation of these parameters from the experimental measurements
is given in {Supplementary material (Appendix~1)}.}

{When including a dendritic segment, the equivalent impedance 
(circuit shown in Fig.~\ref{principle}C, right)} is given by: \b{
\begin{equation}
Z_{eq}^{3,4}(\omega) = Z_i+ \frac{(Z_s + Z_{e}) Z_d}{Z_s + Z_{e} + Z_d} ~ ,
\end{equation}} 
where  \b{$Z_s = \frac{R_s}{1 + i\omega\tau_m}$} is the impedance
of the somatic membrane, \b{$Z_{eq}^{3}(\omega)$} and
\b{$Z_{eq}^{4}(\omega)$} correspond to resistive and diffusive media,
respectively.  Note that in these models, we have considered
\b{$Z_i\approx 0$} (see Fig.~\ref{principle}C) because the
cytoplasm impedance of the soma is negligible compared to the
membrane impedance.

If \b{$i_d^g$} is the current flowing in the dendritic tree, the dendritic 
impedance (as seen by the soma) is: \\
\b{\begin{equation}
Z_d = \frac{V_{intra}}{i_d^g} = \frac{V_{intra}}{V_m} \frac{V_m}{i_d^g} ~.
\end{equation}}
Taking into account \b{$V_{intra} = V_m + V_e$}, we obtain:
\b{
$$
\frac{V_{intra}}{V_m} = 1+ \frac{V_e}{V_m}= 1+ \frac{Z_e}{Z_s}
$$}
because the conservation law for the generalized current implies 
\b{$V_e=Z_e (i^g-i_d^g)$} and \b{$V_m=Z_s (i^g-i_d^g)$}. Note that these 
equalities would not make sense with the free-charge current, because
the variations of \b{$V_m$} may imply charge accumulation around the membrane (dendrite and soma), and thus there is no guarantee of conservation of the free-charge current.

The second part of the fraction represents the input impedance of the 
dendrite \b{$Z_{in}$}, which is given by:
\b{\begin{equation}
\frac{V_m}{i_d^g} = Z_{in} = \frac{\bar{z}_i}{\kappa_\lambda}~ coth(\kappa_\lambda l_d)
\end{equation}}
{where \b{$\kappa_\lambda$} is the cable parameter of the dendrite.} Thus:
\b{
\begin{equation}
Z_d = \left( 1 + \frac{Z_e }{Z_s} \right) \frac{\bar{z}_i}{\kappa_\lambda}coth(\kappa_\lambda.l_d)
\end{equation}}

Note that the values of parameters (\b{$\kappa_{\lambda}$} and
\b{$z_e^{(m)}$} ) in the models considered above correspond to an
open-circuit configuration which corresponds to the present
experiments; the cable equation for the open-circuit configuration,
with an arbitrarily complex extracellular medium, was given
previously~\cite{BedDes2013}.

The intermediate formulas and variables for each model are listed in
Table~\ref{formulas}.

\subsubsection*{Statistics on population data}

Different models call for different sets of parameters. For example,
the membrane resistance and capacitance of a resistive model are
similar but not identical to their counterparts in a model that
features a Warburg impedance (e.g., some of the frequency-filtering
properties can come from this supplementary impedance). Thus, for a
single neuron, we allowed membrane resistance and capacitance to
differ across models.

In order to determine which model was best to describe experimental
data, we used the following classical procedure.

\textbf{1.} For each cell and each model, we computed the residual sum
of squares (RSS) between the experimental curves (\b{$y_{exp}$}) and
theoretical curves (\b{$y_{th}$}): \b{$$ RSS_{cell} =
  \sum_{frequencies} (y_{exp} - y_{th})^2 $$} For each cell, we
normalized this distance by the distance obtained by the best fit. The
distance between experimental and theoretical curves for a given model
was thus: \b{$$RSS_{model} = \sum_{cells} RSS$$}

\textbf{2.} We want to compare the RSS across models. A raw comparison
would be unfair, as a model with more parameters is more capable of
fitting a given data set. Chosing for reference the diffusive model
with dendrites (DD), we thus formed for every other model M the null
hypothesis: ``The model M explains the observed curves. If the model
DD has smaller RSS, it is only because it has more parameters than the
model M.'' 

\textbf{3.} We chose \b{$\alpha < 0.05 $} as threshold for rejecting $H_0$.

\textbf{4.} We used the extra sum of squares F-test, which is able to
account for the discrepancy in degrees of freedom across models.  We
computed the parameter \b{F}, following an F-distribution under $H_0$:
\b{$$ F = \frac{RSS_M - RSS_{DD}}{RSS_{DD}} \frac{DF_{DD}}{DF_M -
    DF_{DD}}$$} where DF is the degrees of freedom (number of cells -
number of parameters) of a given model.

\textbf{5.} The F cumulative distribution function (fcdf) allows us to
compute the likelihood of $H_0$ (Fig.~\ref{modelstat}): \b{$$p = 1 -
  fcdf(F)$$}

If p < \b{$\alpha$}, we reject the null hypothesis: the diffusive model
with dendrites is significantly better than the other model, and this
can not be explained by the surplus of parameters alone.

\subsection*{Computational models {using the measured
    impedances}}

Two types of models were used to test possible consequences of the
measurements.  First, we used a model of the genesis of the
extracellular potential.  To this end, a current waveform
corresponding to the total membrane current generated \b{$I_{AP}(t)$}
by an action potential was computed from the Hodgkin-Huxley model in
the NEURON simulation environment~\cite{Hines97}.  This current
waveform was used as a current source to calculate the extracellular
potential, using a formalism that is valid for any extracellular
impedance.  We calculated the extracellular potential by using the
impedance measurements made in the present paper. According to
\ref{Zwarburg}, we have: \b{
\begin{equation}
Z_e(\omega ) =\frac{A + iB}{1 + \sqrt{i \ \omega / \omega_W}} ~ ,
\end{equation}}
where $A = 151 \times 10^6 \Omega$, $B = 2.54 \times 10^6 \Omega$ 
and $\omega_W = 335~rad.s^{-1}$ for a distance of a few \b{$\mu m$}. 
The extracellular potential \b{$V(t)$} was calculated using the
convolution:
\b{
\begin{equation}
V(t) =  \int_{-\infty}^{+\infty} \bar{Z}_e(t-\tau) I_{AP}(\tau) d\tau ~ ,
\end{equation}}
where \b{$\bar{Z}_e(t)$} is the inverse Fourier transform of 
\b{$Z_e(\omega)$}.

Second, we simulated a ball-and-stick model using the generalized
cable equation~\cite{BedDes2013} {(see Eq.~\ref{gen}).}

A {zero-mean Gaussian white-noise} current waveform was injected
into the dendrite, and the generalized cable was used to compute the
membrane potential in dendrites and in the soma (see details of the
methods in ref.~\cite{BedDes2013}).


\section*{Results}

We start by outlining the measurement paradigm and the notion of
global intracellular impedance, then we successively describe the
results obtained {\it in vitro} and {\it in vivo}.
Finally, we illustrate consequences of
these findings on two fundamental properties, the genesis of
extracellular potentials, and the voltage attenuation along neuronal
dendrites.

\subsection*{The global intracellular impedance}

Here, we explore the hypothesis that the extracellular impedance is
fundamentally different if measured in natural conditions where the
interface between the neuronal membrane and extracellular medium is
respected, compared to metal electrodes, which rely on an artificial
metal-medium interface.  In natural conditions, the neuronal
membrane's interface with the medium involves the opening/closing of
ion channels, ionic concentration differences and ionic diffusion,
whereas metal electrodes involve a different type of ion exchange with
the medium, which consists of a chemical reaction between the metal
and the ions in the medium.  To measure the impedance in natural
conditions, it is therefore necessary to use an intact neuron as the
interface with the medium, to respect the correct ionic exchange
conditions. To do this, we performed whole-cell patch-clamp
recordings, using neurons as natural current sources in the
surrounding medium.

This measurement paradigm is illustrated in Fig.~\ref{principle}A-B
and consists of a whole-cell patch-clamp recording coupled to a
micropipette measuring the potential in the extracellular medium close
to the recorded neuron, {\it in vitro} (Fig.~\ref{principle}A) or {\it
  in vivo} (Fig.~\ref{principle}B).  In all cases, the recorded neuron
is driven by current injection and serves as a natural current source
in the medium.  In this configuration, relating the signals of the two
electrodes gives a direct access to the extracellular impedance, as
shown by the equivalent circuits (Fig.~\ref{principle}C).

According to this equivalent circuit, we have:
\b{
\begin{equation}
	V_{intra} = Z_{eq}. i^g  ~ ,
	\label{equiv}
\end{equation}}
where \b{$i^{g}$} is the generalized current injected by the
patch-clamp electrode.  The use of the generalized current is
essential here because it is the only current that is conserved if the
media have arbitrarily complex impedances~\cite{BedDes2013}, and
therefore abides to Kirchhoff's current laws in this general case.
{Note that} the current provided by the current generator is also
a generalized current because capacitive or non-resistive effects are
negligible in modern generators.

In contrast, the conservation of the classic ``free-charge current''
would apply only with resistive impedances.  The previous equation
gives, for the left circuit (single-compartment cell):
\b{
\begin{equation}
	Z_{eq}(\omega) = Z_i+ \frac{R_m}{1+i\omega \tau_m} + Z_{e} ~ , 
        \label{Z_eq_nodend}
\end{equation}} 
and for the right circuit (cell consisting of a soma and an equivalent dendrite):
\b{
\begin{equation}
	Z_{eq}(\omega) = Z_i+ \frac{(Z_s + Z_{e} ) Z_d}{Z_s + Z_{e} + Z_d} ~ , 
        \label{Z_eq_dend}
\end{equation}} 
where \b{$Z_{eq}$} is the equivalent impedance of each of the two
circuits, \b{$Z_i$} and \b{$Z_e$} are the macroscopic impedances of
the cytosol and the extracellular medium, respectively. \b{$R_m$} is
the global input resistance of the cell and \b{$\tau_m$} is the global
membrane time constant (left circuit); \b{$Z_s = \frac{R_m}{1+i\omega
    \tau_m}$} is the impedance of the soma membrane (right circuit),
\b{$Z_d$} is the input impedance of an equivalent dendrite, as seen by
the soma, including the extracellular medium surrounding it.

In the standard model, \b{$Z_i$} and \b{$Z_e$} are usually modeled by
a lumped and resistive impedance.  In the following, \b{$Z_{eq}$} will
be called the ``global intracellular impedance'' of the circuit,
because in such a recording configuration the neuron acts as a current
source in the brain tissue.  It represents the impedance of the system
as seen by the intracellular side of the neuron.

\b{$Z_{e}$} is the impedance of the extracellular medium as seen by
the neuron (extracellular component of the intracellularly-measured
global impedance).  We will test here if the latter impedance is
negligible or constant, as usually assumed.  We will consider
resistive and diffusive versions of this impedance and check which one
better fits the data.

\b{$Z_i$} includes the impedance of the interface between the tip of
the electrode and the intracellular medium.  This interface will be
different in whole-cell patch or sharp-electrode recording
configurations, because of the location and impedance of the pipettes
themselves.  So, the interpretation of the measured impedances may be
different in sharp-electrode and whole-cell recordings, and we indeed
have observed such differences (not shown here). In particular, we
made sure that the interface of the silver-silver chloride electrodes,
used for patching and as a reference, does not contribute
significantly to the equivalent impedance: it is negligible compared
to other impedances in the circuit and has little frequency dependence
between 1 and 1000 Hz (see Fig.~\ref{Z_electrode}).

\begin{figure}
    \begin{center}
      \includegraphics[width=0.9\columnwidth]{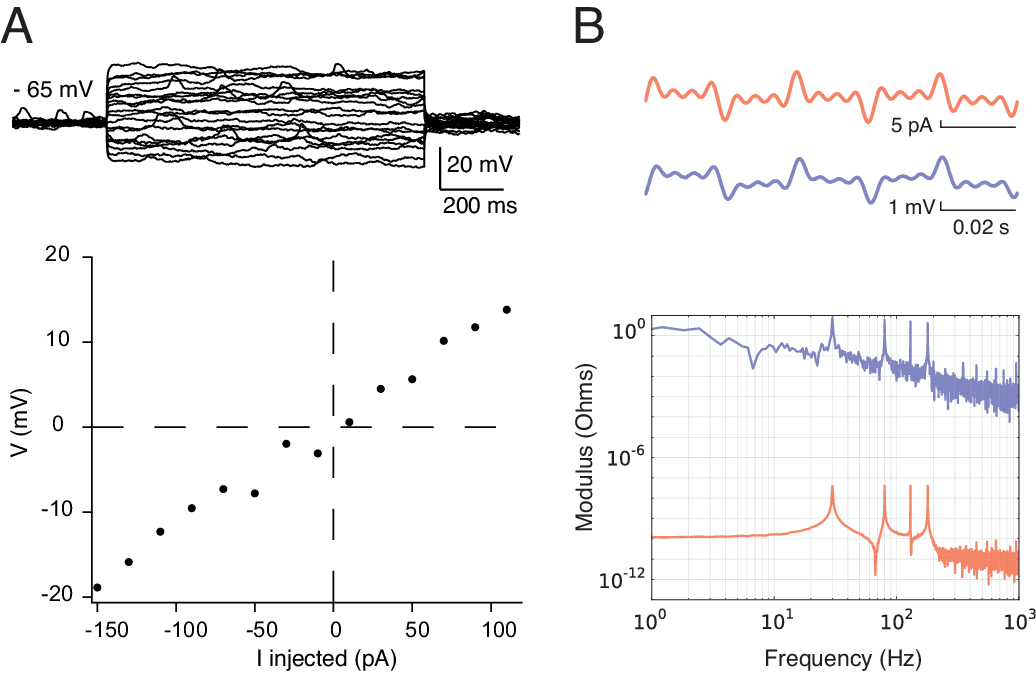}
    \end{center} 

    \caption{{Linearity in temporal and frequency space.  {\bf
          A}: injection of hyperpolarizing and depolarizing pulses in
        a cortical neuron patched {\it in vivo}, around resting
        membrane potential.  The V-I relation indicated is roughly
        linear in this subthreshold range.  {\bf B}: Sine waves of
        current with four different frequencies were injected
        simultaneously in a patched neuron {\it in vivo} (top, red
        curve).  The voltage was recorded (top, blue curve).  The
        modulus of the Fourier Transform of both signals is shown
        here, as a function of frequency (Hz): current in red (A),
        voltage in blue (V).  The frequencies corresponding to the
        peaks on both signals illustrate the linearity in frequency of
        the system, here tested between 1 and 1000 Hz.}}

\label{linearity} 
\end{figure}

In order to measure \b{$Z_{eq}$}, we injected a current
intracellularly, and measured the intracellular potential
\b{$V_{intra}$} with respect to a reference.  Ideally, this reference
is a micropipette in the extracellular medium (\b{V$_{LFP}$}; see
scheme in Fig.~\ref{principle}A).  In the {\it in vivo} experiments,
the reference was a silver electrode inserted in the contralateral
somatosensory cortex as in Fig.~\ref{principle}B.  For subthreshold
currents, the system can be considered linear in frequency: injecting
current at an arbitrary frequency yields voltage variations only at
that frequency {(see {Appendix~2 in Supplementary Material})}.  {This
  linearity of the system is illustrated in Fig.~\ref{linearity}.
  First, for the subthreshold range of $V_m$ considered, the membrane
  V-I relation is linear (Fig.~\ref{linearity}A).  Second, a
  combination of four sine-wave currents of fixed frequency generate
  $V_m$ variations that have strictly the same frequencies
  (Fig.~\ref{linearity}B),} showing that the linear approximation is
valid: sine waves appear as sharp {spectral lines} in Fourier space,
two orders of magnitude above baseline.  Thus, there seems to be no
significant impact of non-linear membrane ion channels on the
recordings.  {Indeed, no action potentials were present in the
  recordings analyzed here.}

We used two protocols of current injection, either injection of a
series of sinusoidal currents of different fixed frequencies (6 to 926
Hz), or injection of a broad-band (white noise) current, with a flat
spectrum between 1 and 10~kHz.  In this case, several instances of the
same noisy current trace were injected in the cell (``frozen noise''
protocol), which allowed us to take average and increase the
signal-to-noise ratio of the measurements.  This is particularly
useful {\it in vivo}, to limit the contamination of the measurement by
spontaneous synaptic activity, which can be very strong {\it in vivo}.

\subsection*{\emph{In vitro} measurements of the global
  intracellular impedance}

\begin{figure}[!ht]

   \begin{center}
      \includegraphics[width=0.9\columnwidth]{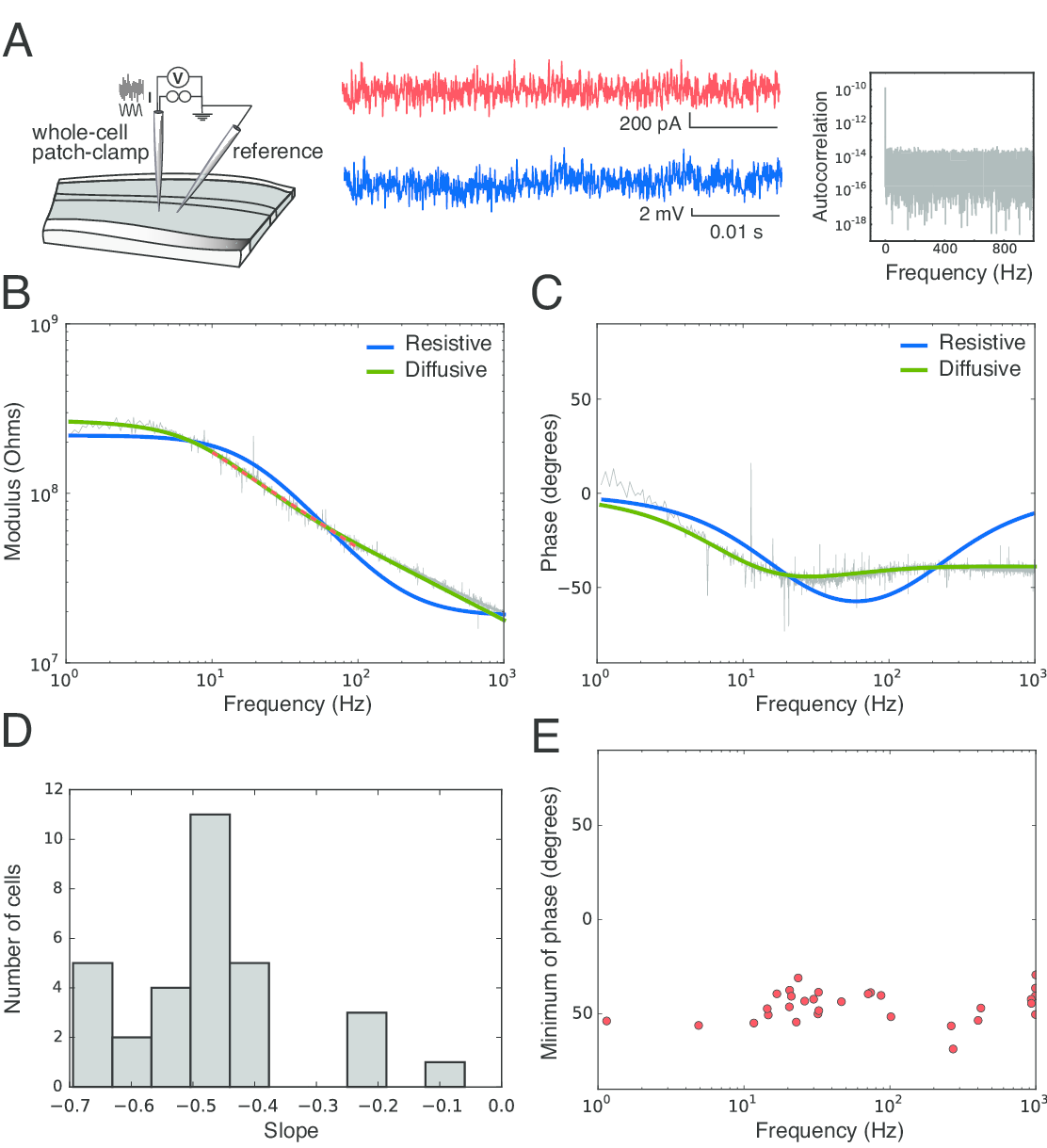}
   \end{center} 

   \caption{\label{invitro} Global intracellular impedance of cortical
     neurons recorded {\it in vitro} (current clamp). {\bf A}:
     {Left:} Scheme of the experimental setup; {Middle:}
     example signals; A gaussian white noise current signal (red,
     amplitude: $\pm$100 pA) was injected repeatedly in patched
     neurons while recording the intracellular potential (blue).  For
     each 20 seconds period, we computed the impedance as the ratio of
     measured voltage on injected current (\b{$Z_{mes} =
       \frac{V_{intra}}{i_{inj}}$}) and averaged on 30 periods.
     Acquisition was performed at 20 kHz. {Right: Autocorrelation
       of the injected white noise current in Fourier space.}
     \textbf{B}: modulus of \b{$Z_{eq}$} (log scale) represented as a
     function of frequency (\b{$log_{10}(f)$}), for white noise
     current injection.  \textbf{C}: Fourier phase of \b{$Z_{eq}$}, in
     the same experiment.  The different curves shows the best fit of
     two models to the experimental data (blue: resistive, green:
     diffusive).  \textbf{Parameters of the models}: i) Resistive:
     $\b{R_m} =~200 M\Omega$, $\b{C_m} = 45~pF$, $\b{R_e} =
     19~M\Omega$; ii) Diffusive: $\b{R_m} = 180~M\Omega$, $\b{C_m}
     =~110 pF$, $\b{A} =~99 M\Omega$, $\b{B} = 3.8~M\Omega$, $\b{f_W}
     = 36~Hz$. $\b{A}$, $\b{B}$ and $\b{f_W}$ are parameters of the
     diffusive impedance scaling respectively its amplitude, phase,
     and cutoff frequency.  \textbf{D}: Distribution of the slopes of
     \b{$Z_{eq}$} fitted between 20 and 200 Hz (linear fit, red dashed
     lines on the figures).  \textbf{E}: Coordinates of the minima of
     the Fourier phases for each cell.}

\end{figure}

Measurements were first performed {\it in vitro} by using an
experimental setup consisting of a whole-cell patch-clamp recording of
a neuron, together with an extracellular recording in the nearby
tissue in the cortical slice (see Fig.~\ref{invitro}A).  Using this
recording configuration, we computed the global intracellular
impedance (\b{$Z_{eq}$} in Eq.~\ref{equiv}; see Materials and Methods)
by using either white-noise current injection, or injected sinusoidal
currents.  The results of a representative cell ($N=31$) is shown in
Fig.~\ref{invitro}B-C.  Both the modulus amplitude and Fourier phase
of the impedance are represented. The colored curves in
Fig.~\ref{invitro}B-C show the best fits of different models to the
experimental data.  One can see that the purely resistive model (RC
membrane + resistive extracellular medium, blue curves) {does not} 
capture the data.  We read from equation \ref{eqresistive} that
\b{$\lvert Z_{eq}(\omega) \rvert$} scales as \b{$1/\omega$} in the
resistive model, which corresponds to a slope of -1, while the
experimental modulus yields a slope of $-0.5 \pm 0.1$ (Fig.
\ref{invitro}D). The resistive model has a phase similar to
\b{arctan($k\omega$)} with a minimum of about -90 degrees at high
frequencies, which contrasts with the -50 degrees observed in the data
(Fig.~\ref{invitro}E).

\begin{figure}[!ht]
  \begin{center}
     \includegraphics[width=0.7\columnwidth]{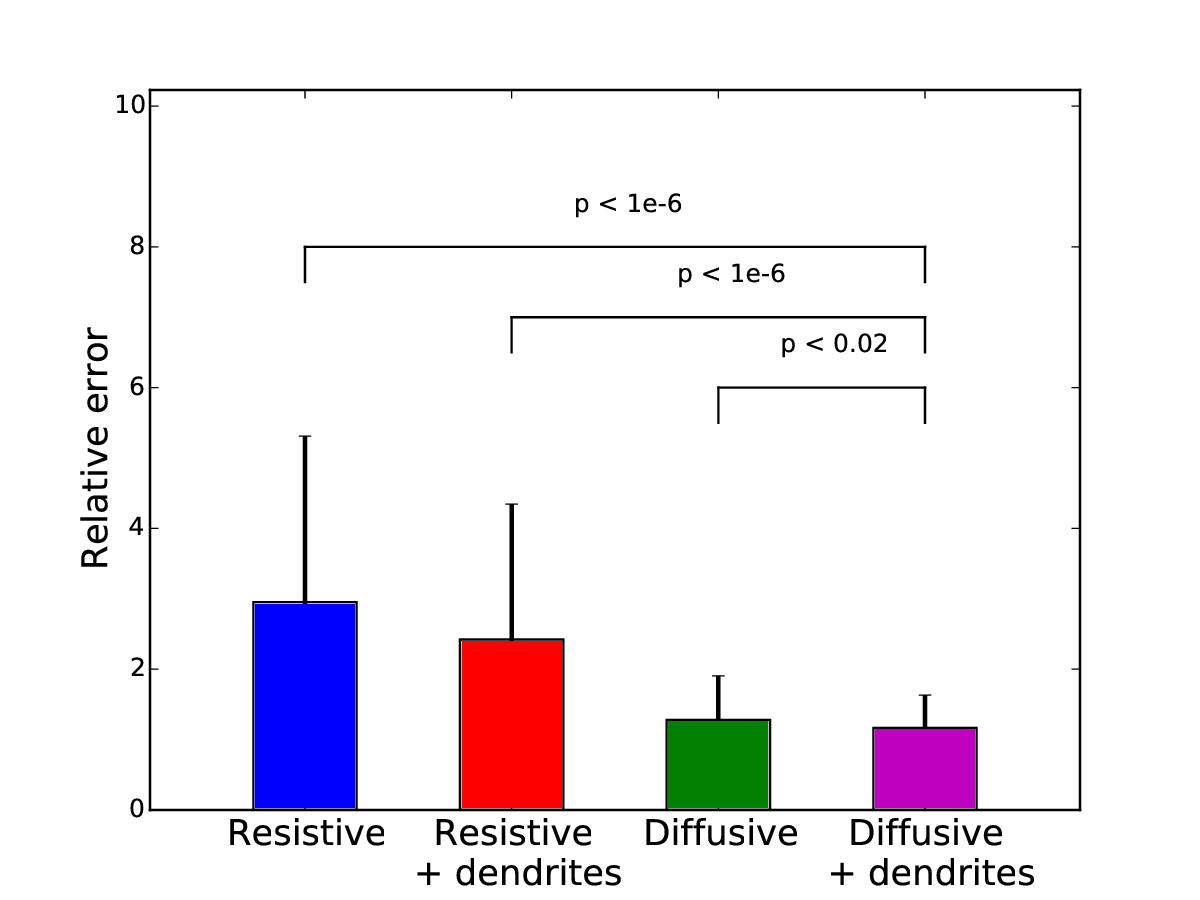}
  \end{center}

  \caption{Average fitting error of the four models.  The figure shows
    the average goodness of fit for 4 different models investigated:
    {\it Resistive}, {\it Resistive with Dendrite}, {\it Diffusive},
    and {\it Diffusive with Dendrite}. The error bars are
      standard deviations of fitting errors. The testing variable
    used to compare the quality of these fits takes into account their
    different number of variables. Of all models, the diffusive models
    give the smallest error in all cases tested. The best fit is
    provided by the diffusive model with dendrite, although only
    marginally better than the diffusive model in a single-compartment
    model.}

\label{modelstat}
\end{figure}

So frequency dependence is clearly different from that predicted by
the RC-circuit membrane model.  The best fits of a model taking into
account ionic diffusion~\cite{BedDes2011} can account significantly
better for most of this frequency dependance in different cells (green
curves; see also Fig.~\ref{modelstat}).  In particular, the
\b{$1/\sqrt{\omega}$} frequency scaling predicted by the diffusive
model (equation \ref{eqdiffusive}), corresponds to the actually
observed -0.5 slope of the modulus.  The phase modulations can also be
remarkably well captured by the diffusive model.

\begin{figure}[!ht]

  \vspace{4cm}

  \begin{center}
   \includegraphics[width=\columnwidth]{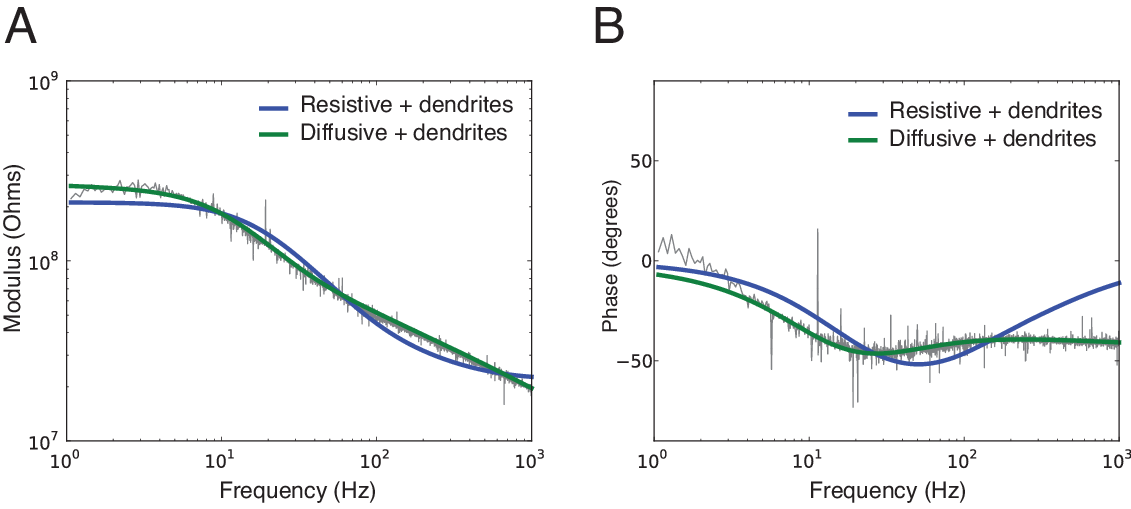}
  \end{center}  

  \caption{\label{dendrites} Dendritic contribution to the global
    intracellular impedance of a cortical neuron recorded {\it in
      vitro} (current clamp).  Same arrangement of panels as in
    fig.~\ref{invitro}, except that the different curves show the best
    fit of two models to the experimental data (blue: resistive,
    green: diffusive), both models included an equivalent dendritic
    compartment.  \textbf{Parameters of the models}: i)
    Resistive: $\b{R_s} =~240 M\Omega$, $\b{C_s} = 37~pF$, $\b{R_e} =
    21~M\Omega$, $\b{l_{dend} = 390~\mu m}$; ii) Diffusive: $\b{R_s} =
    150~M\Omega$, $\b{C_s} =~89 pF$, $\b{A} =~130 M\Omega$, $\b{B} =
    -12~M\Omega$, $\b{f_W} = 30~Hz$, $\b{l_{dend} = 12~\mu m}$.
    $\b{A}$, $\b{B}$ and $\b{f_W}$ are parameters of the diffusive
    impedance scaling respectively its amplitude, phase, and cutoff
    frequency.}

\end{figure}

We also tested the possible influence of dendrites, by including an
equivalent dendritic compartment in the circuit
(Fig.~\ref{principle}C, right).  This addition could not rescue the
resistive model, which was still unable to match the observations
(Fig.~\ref{dendrites}). {We have considered variations of
  dendritic parameters, including very long dendritic segments,
  different axial and leak conductances, and did not see any
  significant improvement by the addition of dendrites.} In the
diffusive model, taking the dendrites in account only enhanced
marginally the agreement between experimental and theoretical curves.
Statistical analysis showed that the improvement in quality from the
resistive to the diffusive model was significant, and not only due to
a higher number of parameters.  Furthermore, the apparent smaller
number of parameters in resistive models can come from hidden
assumptions, such as homogeneity and low resistivity of the
extracellular medium.

\begin{figure}[!ht]

  \begin{center}
     \includegraphics[width=0.9\columnwidth]{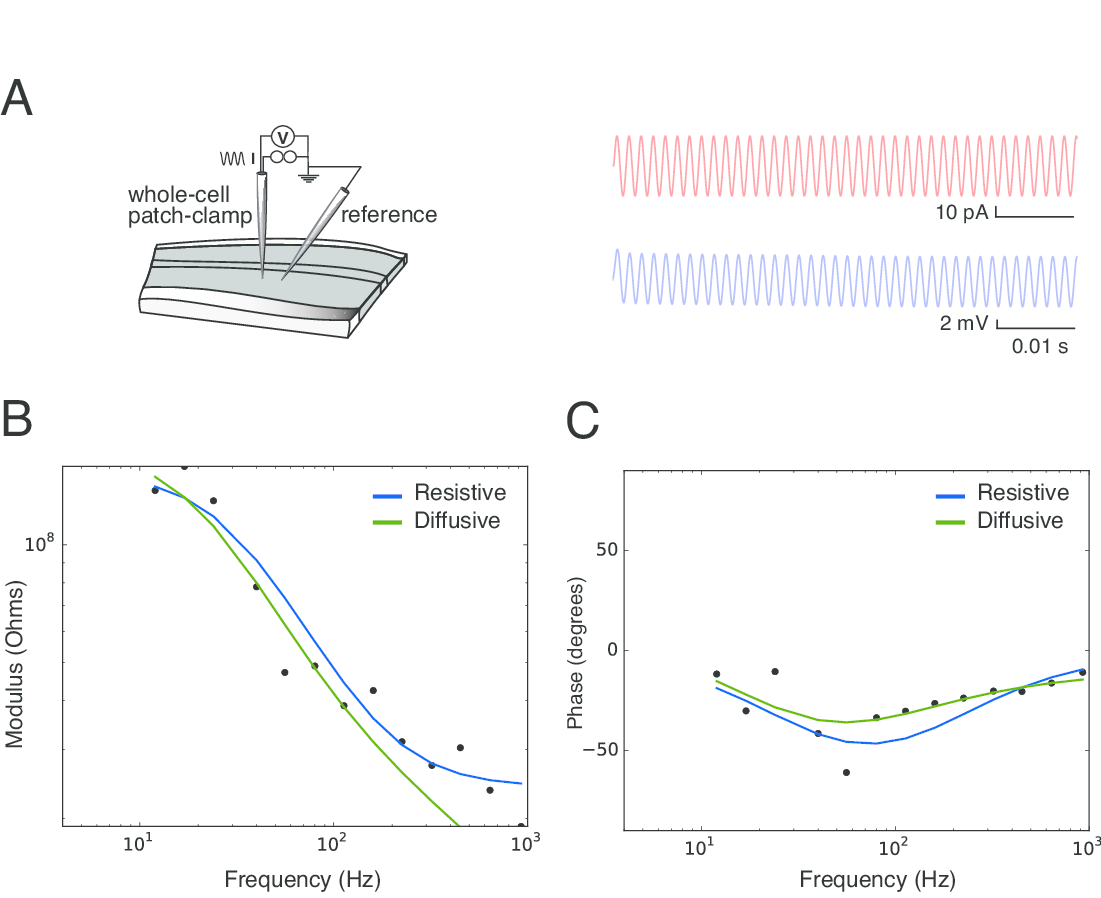}
  \end{center}

    \caption{Global intracellular impedance measurements of a
      {striatal neuron recorded {\it in vitro} (current clamp)
        and stimulated with sinusoid inputs from 6 to 926 Hz}. {\bf
        A}: sine waves of current (top, red; minimum 25 samples and 20
      cycles per sample) were injected in patched neurons while
      recording the intracellular potential (top, in blue; 16.7~kHz
      sampling).  {\bf B}: modulus of \b{$Z_{eq}$} (log scale)
      represented as a function of frequency (\b{$log_{10}(f)$}), for
      sine wave current injection (600 pA amplitude).  {\bf C}:
      Fourier phase of \b{$Z_{eq}$} in the same experiment.
      \textbf{Parameters of the models}: i) Resistive: $\b{R_m} =
      120~M\Omega$, $\b{C_m} =~46 pF$, $\b{R_e} = 28~ M\Omega$; ii)
      Diffusive: $\b{R_m} = 130~ M\Omega$, $\b{C_m} = 59~pF$, $\b{A} =
      58 M\Omega$, $\b{B} = 43~M\Omega$, $\b{f_W} = 50~Hz$.
      Representative of $N=18$ cells.}

\label{invitrosin} 
\end{figure}

These results were replicated {in striatal neurons} using purely
sinusoidal input currents from 6 to 926 Hz (see Fig.~\ref{invitrosin},
{$N=18$}), {thereby confirming that the global shape of the
  global cell impedance does not depend on the stimulation protocol.}
In addition, we tested a capacitive (RC) model of the extracellular
space, but this model {did not} account for the modulus and the
phase modulations ({it was the worst fit of all models tested} --
not shown).

We {also} checked whether the quality of seals could affect the
global intracellular impedance measurements.  Indeed, if the cell
membrane is bypassed, the impedance is not measured anymore through
the natural interface of a neuron membrane. The average of neurons
with good seals (> 1 G$\Omega$) yields a slope of $-0.5 \pm 0.1$ (see
fig.  \ref{invitro}D); in comparison, cells with extremely poor seals
(e.g.  200 M$\Omega$, not included in the data shown here) yielded a
flatter impedance, with a slope between 0 and -0.3.  This can be
easily {explained by replacing \b{$Z_{RC}$}} by a resistance in the 
expression of \b{$Z_{eq}$}.

{Finally, we} checked whether part of the observed frequency
dependence could be due to the recording pipette.  We found that the
frequency dependence of the patching pipette and silver-silver
chloride electrode is negligible in ACSF (Fig.~\ref{Z_electrode}).
These measurements show that the observed frequency dependence of the
impedance cannnot be attributed to the silver electrode interface, and
probably stems from the properties of the extracellular medium.

\subsection*{Global intracellular impedance \emph{in vivo}}

\begin{figure}[!ht]

  \begin{center}
     \includegraphics[width=0.9\columnwidth]{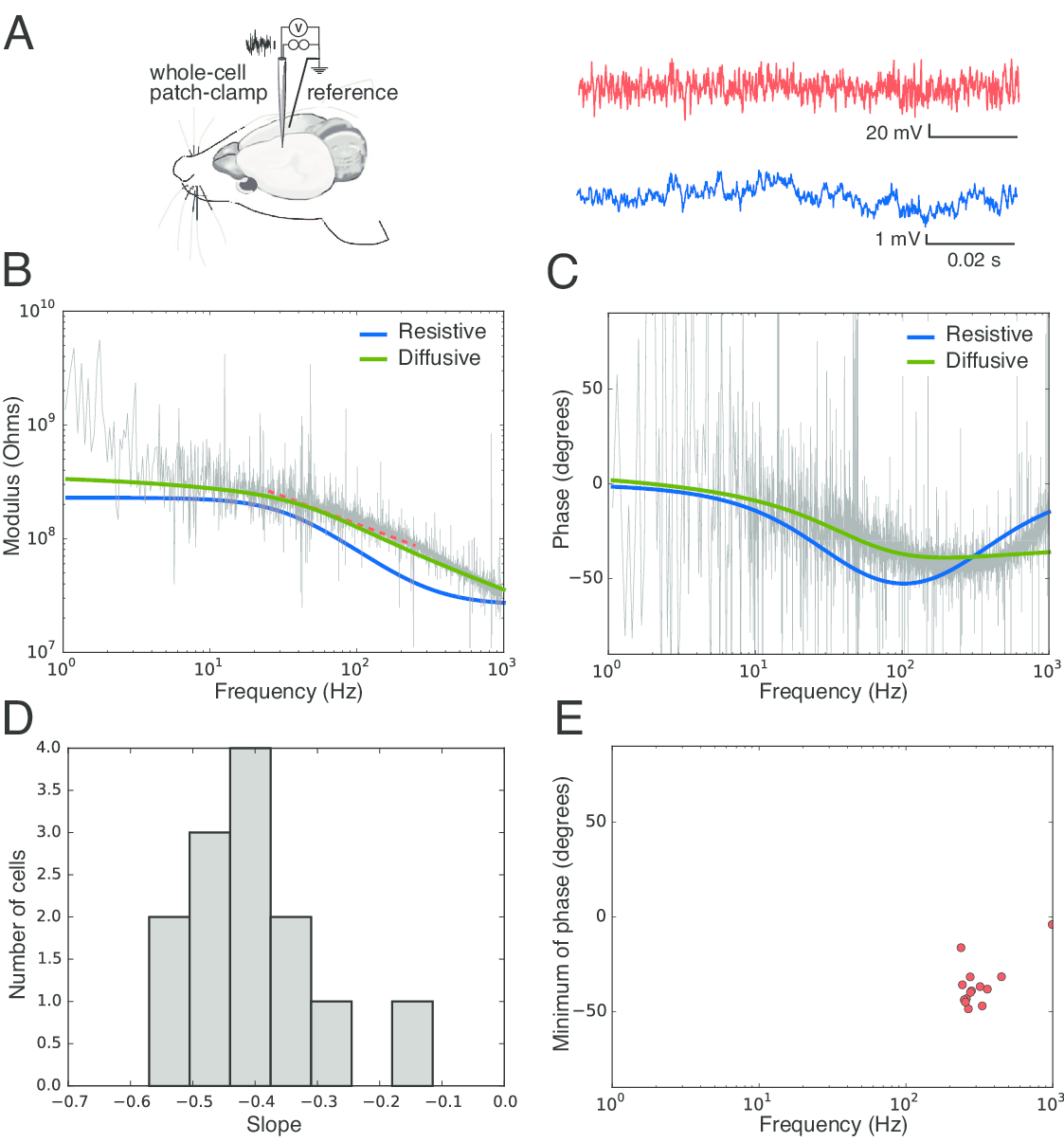}
  \end{center}
	
  \caption{\label{invivo} Global intracellular impedance of cortical
    neurons recorded {\it in vivo} {in current-clamp}.  {\bf A}:
    Scheme of the recording configuration, with a similar
    representation as in Fig.~\ref{invitro}A, but performed in rat
    somatosensory cortex {\it in vivo} (with a contralateral reference
    electrode), using white noise current injection.  For these
    experiments, we injected 20 times the same gaussian white noise
    current trace (top, in red), recorded at 20 kHz and averaged the
    intracellular potential (top, in blue).  We calculated the
    impedance seen by the neuron.  {\bf B}: Modulus of the impedance
    obtained, as a function of frequency.  {\bf C}: Phase of the
    impedance. \textbf{Parameters of the models}: i) Resistive:
    $\b{R_m} = 230~M\Omega$, $\b{C_m} = 28~pF$, $\b{R_e} =
    29~M\Omega$; ii) Diffusive: $\b{R_m} = 100~M\Omega$, $\b{C_m} =~21
    pF$, $\b{A} = 250~M\Omega$, $\b{B} = 24~M\Omega$, $\b{f_W} =
    20~Hz$ Representative of {$N=18$} cells.  \textbf{D}:
    Distribution of the slopes of \b{$Z_{eq}$} fitted between 25 and
    250 Hz (linear fit, red dashed lines on the figures). \textbf{E}:
    Coordinates of the minima of the Fourier phases for each cell.}

\end{figure}

In a second set of experiments, the measurements were performed {\it
  in vivo} with whole-cell patch-clamp recordings {of pyramidal 
cells of the  somatosensory cortex, layer V} (see scheme in
Fig.~\ref{invivo}A and details in Materials and Methods).  Similarly
to Fig.~\ref{invitro}, the modulus and phase of the impedance were
estimated by white noise current injection (Fig.~\ref{invivo}B-C).
Although the data display a high degree of noise (due to spontaneous
synaptic inputs {\it in vivo}), they were in qualitative agreement
with {\it in vitro} results on {$N=18$} cells.  The resistive model did
not capture the modulus amplitude, nor the phase of the global
intracellular impedance of the neuron.  The diffusive model was able
to capture the essential variations, both in amplitude (modulus) and
phase domain.  Similar to {\it in vitro} measurements, the modulus
yields a slope of $-0.4 \pm 0.1$ (Fig.~\ref{invivo}D), and a minimum
phase around -50 degrees (Fig.~\ref{invivo}E), which significantly
deviate from predictions of a resistive model.

As in the {\it in vitro} experiments, the addition
of an equivalent dendritic compartment did not improve these
differences.  The resistive model with dendrites was also unable to
account for the measurements, while the diffusive model provided
acceptable fits to the data.

{One must keep in mind that the {\it in vivo} measurements were
  made in the presence of low-frequency spontaneous activity, typical
  of anesthetized states.  This ``synaptic bombardment'' probably
  explains the mismatch of all impedance models at low frequencies
  {\it in vivo}.  Such a mismatch was not present {\it in vitro}.}

\subsection*{Possible consequences of these measurements}

\begin{figure}

  \begin{center}
    \includegraphics[width=\columnwidth]{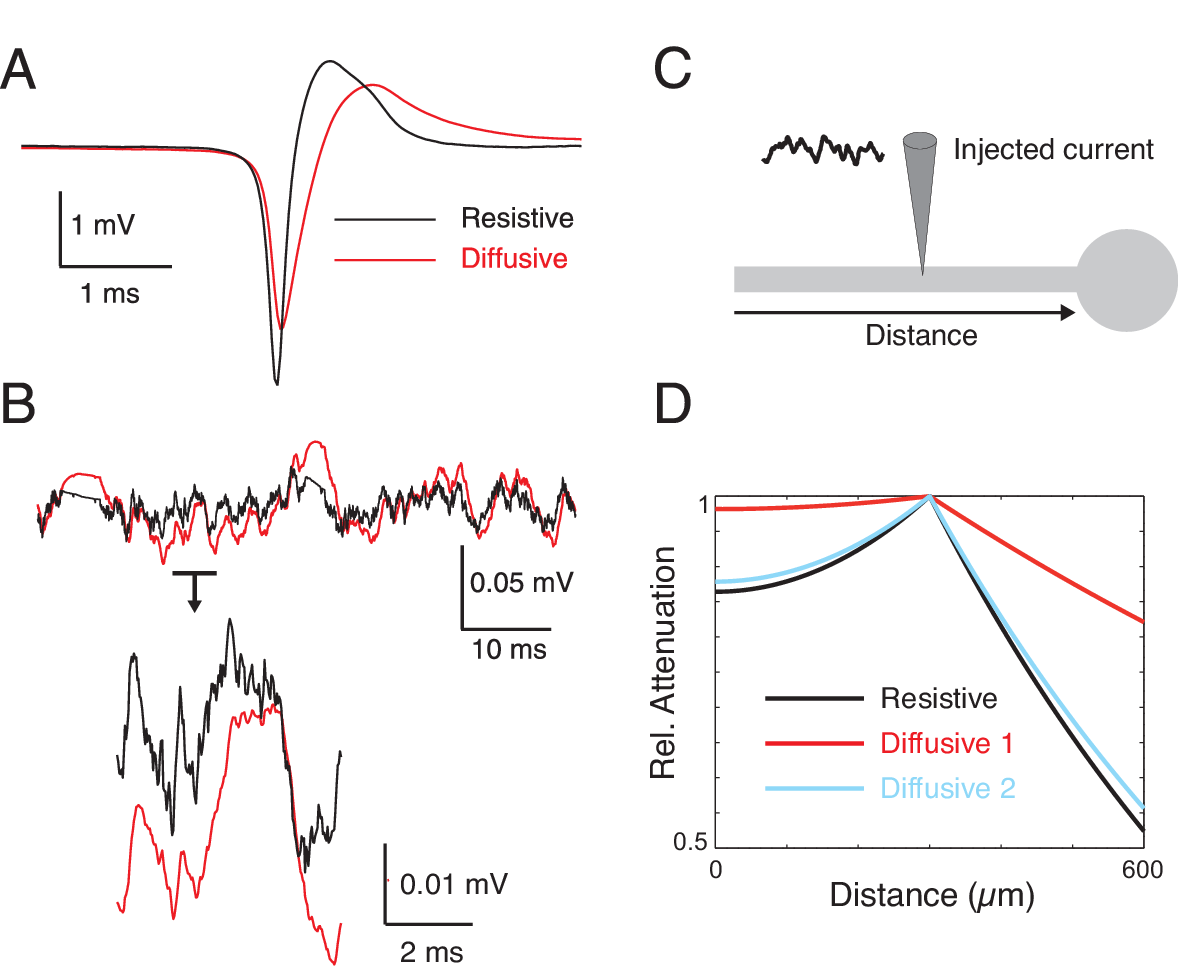}
  \end{center}  	

  \caption{Consequences of the diffusive nature of the extracellular
    medium.  {\bf A}. Simulation of LFP in the extracellular medium
    {(10~$\mu$m from the soma)} following injection of current
    following a spike waveform.  The panel compares the extracellular
    spike obtained for a resistive medium (black), compared to a
    diffusive medium (red), where the filtering is evident.  {\bf B}.
    Same simulation as in A, but using injection of the combined
    current of noisy excitatory and inhibitory (subthreshold) inputs.
    In this case, the LFP obtained in a resistive and diffusive medium
    are also differentially filtered.  {\bf C}. Scheme of a
    ball-and-stick neuron model where a noisy current waveform was
    injected in the middle of the dendrite.  {\bf D}. Relative voltage
    attenuation profile obtained (at 5Hz) when the neuron is simulated
    in resistive (black) or diffusive (red, {blue}) media.  {Two
      diffusive configurations were simulated, {\it Diffusive~1}:
      diffusive intracellular and extracellular media (red curves);
      {\it Diffusive~2}: diffusive extracellular medium with resistive
      intracellular medium (blue curves).}  Using diffusive media
    results in a reduced voltage attenuation.  In all cases, the
    resistive or diffusive media were simulated using the best fits to
    the impedances measured {\it in vitro}.}

\label{conseq}
\end{figure}

Finally, to evaluate possible consequences of these measurements, we
have considered two situations where the extracellular impedance can
have strong consequences.  A first consequence is the fact that the
diffusive nature of the medium will necessarily impose frequency
filtering properties on extracellular potentials, which affects
measurements made with extracellular electrodes. To illustrate this
point, we simulated extracellular potentials generated by a current
source corresponding to the total membrane current generated by an
action potential (using the Hodgkin-Huxley model).  We then calculated
the extracellular potential at a distance from this current source,
using either a resistive model, or a diffusive model
(Fig.~\ref{conseq}A).  Interestingly, one can see that the
extracellular signature of the spike has a slower time course in
diffusive conditions.  A similar situation was also simulated using
subthreshold noisy excitatory and inhibitory synaptic activity
(Fig.~\ref{conseq}B).  In both cases, the nature of
the medium influences the shape and propagation of the local field
potential (LFP), for both extracellular spikes and LFP resulting from
synaptic activity.

A second possible consequence is on the cable properties of neurons.
This point was illustrated by simulating a ball-and-stick model
subject to injection of a noisy current waveform in the dendritic
cable (Fig.~\ref{conseq}C).  As shown in Fig.~\ref{conseq}D, the
attenuation of the voltage along the dendrite can be drastically
different in a diffusive medium compared to a resistive medium, as
noted previously~\cite{BedDes2013}.  {Including a diffusive
  extracellular medium reduced voltage attenuation
  (Fig.~\ref{conseq}D, blue curve), but this reduction was the
  strongest when both intracellular and extracellular media were
  diffusive (red cuves).} Thus, the nature of the medium will also
influence the shape and propagation of potentials in dendrites.

\section*{Discussion}

We have provided here the first experimental measurements of the
impedance of the extracellular medium, in natural conditions, {\it in
  vitro} and {\it in vivo}.  We found that, not only the estimated
extracellular impedance is higher compared to {traditional}
metal-electrode measurements, but it is also more frequency dependent.
The standard model, considering the medium as resistive, can account
for metal electrode measurements, but not for natural impedance
measurements. In contrast, we found that a diffusive model can account
for most measurements, both in modulus amplitude and in phase.  We
also checked whether the inclusion of dendrites could affect these
conclusions, but it did not qualitatively change these results.

It is noteworthy that the present measurements are made between a
neuron, and a reference electrode in the nearby tissue.  Therefore,
the current presumably flows in the entire tissue, and thus, the
impedance measured can be considered ``macroscopic''.  From the
different experiments realized here, we estimate that {the impedance
  is determined by the region close to the membrane, within distances
  of the order of hundreds of microns in cerebral cortex.}

The apparent inconsistency with the previous metal-electrode
measurements can be resolved by considering that each kind of
electrode has a specific interface and impedance, depending on its
physical nature~\cite{Geddes}. Classical impedance measurement studies
tackle this problem offline with a normalization by a measure in
saline~\cite{Ranck,Gabriel}, online by removing the effect of the
interface using the saturation due to large
currents~\cite{Logothetis}, or by minimizing the
interface~\cite{Robinson,Schwan}. In physiological conditions,
neurons have an electrical interface with the extracellular medium, as
a part of their normal environment. This interface should therefore
not be removed when using neurons to evaluate the impedance of the
extracellular medium, as it is one of the keys to explaining the
electric field produced by an active cell.

\begin{figure}
    \begin{center}
      \includegraphics[width=0.9\columnwidth]{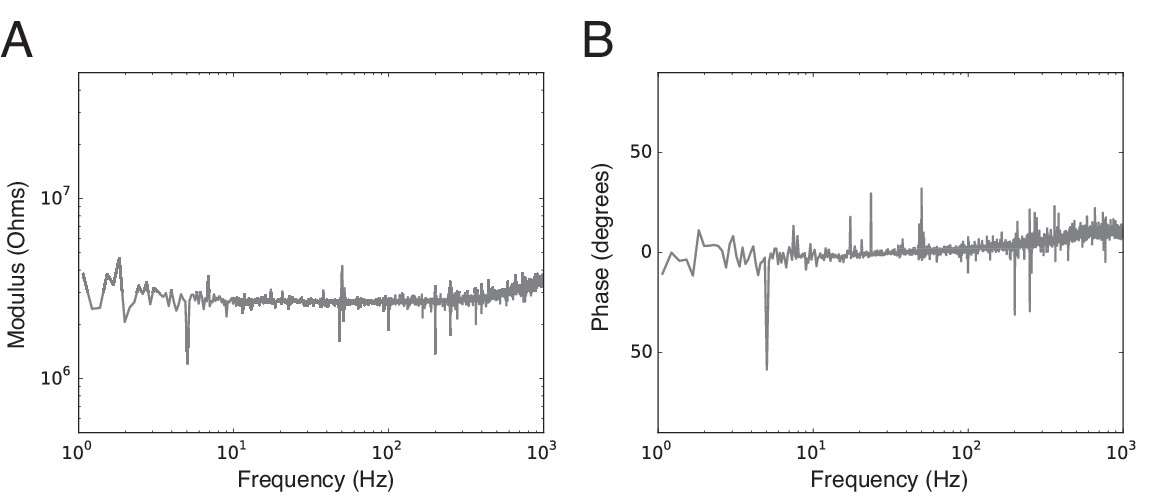}
    \end{center} 

    \caption{Impedance of a silver-silver chloride electrode inserted
      in a typical ($\approx 3~M\Omega$) patch pipette, measured in
      ACSF.  Left: modulus (Ohms), right: phase (degrees).}

\label{Z_electrode} 
\end{figure}

The system presented here deals with the usual problems of electrode
recordings (see ref.~\cite{Robinson}) in unusual ways, which solves
some classical issues but raises new interrogations.  First, the
electrode -- or neuron -- must be standardized. It is remarkable to
see that, despite the considerable cell-to-cell variability in size or
morphology, we obtained here very consistent measurements, with very
similar amplitude and phase profiles from cell to cell. These
measurements can be captured with accuracy with a limited number of
parameters, most of which are well known (\b{$R_m$}, \b{$C_m$}...).
{How the cellular morphology influences these results, and why
  this influence seems so small, constitute interesting subjects for
  future extensions of the present work.}  Second, the spatial scale
concerned by potential measurement and current injection is a grey
area in the litterature (see ref.~\cite{Nunez2005}). We believe that
the scale of a single neuron may be as relevant as the tip of
traditional recording electrodes, of arbitrary size and position in an
inhomogenous medium, which can affect recordings
significantly~\cite{Nelson2013}. Third, the interface of the electrode
and its behavior must be linear and well understood within the
measurement range, which we discussed previously.  Provided the system
is operated with all necessary precautions, linearity is maintained
{(Fig.~\ref{linearity});} the path of the injected currents in
neuron compartments does not seem to be a crucial matter
(Fig.~\ref{Z_eq_dend}) and injected currents splitting between
different forms (free or bound charges, electric flux...) is not a
problem within the generalized current formalism.  Furthermore, the
traditional 4-electrode setup is designed to separate voltage
recording from possible filtering by the interface generated when
injecting current~\cite{Schwan}. In the system presented here, the
silver-silver chloride wire has a very resistive interface
(Fig.~\ref{Z_electrode}) and is negligible with respect to the main,
relevant interface of the recorded neuron.

A possible explanation for the prominent role of ionic diffusion is
that when a neuron acts as a current source, the electric field lines
{will not, in general, match} the complex geometry of the
extracellular medium. The trajectory of ions would thus be affected by
obstacles such as cells and fibers~\cite{Nelson2013}, which would
yield local variations in ionic concentrations.  Ionic diffusion would
therefore exert an important force on ions in the extracellular
medium. A linear approximation of this phenomenon allows one to model
this contribution by a Warburg-type impedance, scaling as
$\frac{1}{\sqrt{\omega}}$.  In addition, ionic diffusion is involved
in membrane potential changes, and participates as well to maintaining
the Debye layer surrounding the membrane~\cite{Hille2001}. Taken
together, these factors could explain why the present measurements are
in such good agreement with the diffusive model.


{Despite this agreement, the participation of diffusive phenomena
  can vary with age and experimental conditions.  As the brain gets
  older, the extracellular volume fraction shrinks, which could make
  ionic diffusion even stronger and thus reinforce the Warburg
  component of the natural impedance.  Furthermore, in vivo tissue may
  be more confined than in vitro, with a similar result on the
  importance of ionic diffusion. One can thus reasonably expect
  \b{$Z_W$} to be stronger in P30 rats in vivo ($N=18$) than in P12-16 mice in
  vitro ($N=31$). Indeed, the components \b{$A$} and \b{$B$} of \b{$Z_W$} are
  significantly stronger \emph{in vivo} in P30 rats than \emph{in
    vitro} in P12-16 mice (comparing the medians $\pm$ the standard 
    error of the mean: resp. $96 \pm 12~M\Omega$ vs $183 \pm 16~M\Omega$
  for \b{$A$} and $6.6 \pm 2.7~M\Omega$ vs $28 \pm 3.7~M\Omega$ for
  \b{$B$}). Thus, the age of the subject and type of recording need to
  be taken into account when using measured values of the natural
  impedance. In particular, it may lead to an underestimation of
  \b{$Z_W$} in this paper, because we mainly focused on \emph{in
    vitro} recordings in young mice; our conclusions on the importance
  of ionic diffusion are thus rather conservative.  It is noteworthy
  that in between these two sets of observations, the Warburg
  frequency remains the same : $43 \pm 3~Hz$ \emph{in vitro} vs $42
  \pm 3~Hz$ \emph{in vivo}.}

Our results do not disqualify the previous measurements, but are
complementary.  We suggest that for all cases where the current
sources are generated by natural conditions (i.e., by neurons), the
global intracellular impedance should be used. This is the case for
example when analyzing the LFP signal, or with Current Source Density
(CSD) analysis.  In cases where a metal electrode is used to inject
current, the metal-electrode impedance would be relevant, for example,
in Deep Brain Stimulation paradigms.

Note that, although the diffusive model accounts very well for the
modulus and phase variations of the global intracellular impedance,
there exists small deviations, in particular at high frequencies. The
latter may be due to a number of phenomena, including variability in
neuron geometry or limitations of the linear approximations used here.
The existence of ``shunts'' due to the liquid around the electrodes is
also not to be excluded.  Further studies should be designed to
identify the contribution of such factors, e.g.  pharmacological
inactivation of nonlinear channels. Two arguments suggest that the
present formalism is satisfactory, the strong reproducibility of
results across 31 recorded neurons \emph{in vitro}, despite intrinsic
biological variability, and the coherence between the diffusive model
and experimental data.

{The exact boundaries of the domain where these results apply are
  still to be determined.  For example, in figure \ref{conseq} we are
  extrapolating into a nonlinear region to make implications about the
  shape of the action potential. We think this extrapolation is
  acceptable because non-linear behavior is mostly happening in the
  highest frequencies, barely overlapping with the LFP frequency range
  (see also Appendix~2 {in Supplementary Material}), but one should be
  aware of that caveat.}

Finally, using computational models, we illustrated consequences of
the medium non-resistivity on extracellular and intracellular
potentials.  A number of fundamental theoretical equations used in
neuroscience, such as CSD analysis~\cite{Mitzdorf}, or neuronal cable
equations~\cite{Rall1962,Rall1995} were originally derived under the
assumption that the extracellular medium is resistive.  If the medium
is non-resistive, these equations are not valid anymore and must be
generalized.  Attempts for such generalizations were proposed recently
for CSD analysis~\cite{BedDes2011} and cable
equations~\cite{BedDes2013}, but they were not constrained by
measurements.  The simulations provided here show that including a
diffusive impe\-dance based on the present measurements has
significant consequences, for both extracellular potentials, and for
the electrotonic properties of neurons.  The shape of the
extracellular spike may be affected by the nature of the medium
(Fig.~\ref{conseq}A), {assuming that one can extrapolate the
  present results to the nonlinear region of the $V_m$.  This shows
  that the sharpness of the extracellular spike may be influenced by
  the properties of the medium,} which constitutes another factor that
could complicate the identification of neurons from spike shape.  The
dendritic attenuation is also reduced in the presence of a diffusive
medium (Fig.~\ref{conseq}D-E), as shown previously~\cite{BedDes2013}.
Extrapolating these results, it seems that the sources estimated by
CSD analysis, {could greatly be affected by the nature of the
  extracellular medium, which constitutes a direct extension of the
  present study.  Similarly, source reconstruction methods from the
  EEG, are also likely to be affected by the nature of the medium, and
  thus, these methods may need to be re-evaluated as well}.

\subsection*{Author contributions}

\noindent JMG, SV, MN, PP and TB performed the {\it in vitro}
experiments.  LV performed the {\it in vivo} experiments.
Analyses were designed by AD and CB and were performed by JMG, CB, MN,
VK and AD. AD, TB and LV cosupervised the work.

\subsection*{Acknowledgments} 

Research supported by the CNRS, {the Paris-Saclay excellence network
  (IDEX)}, INSERM, Coll\`ege de France, Fondation Brou de Lauri\`ere,
Fondation Roger de Spoelberch, French Ministry of Research, the ANR
(Complex-V1 project), the Eiffel excellency grants program, and the
European Community (BrainScales FP7-269921, Magnetrodes FP7-600730 and
the Human Brain Project FP7-604102).  We thank Sylvie Perez for
technical assistance with the in vivo experiments.


\label{Bibliographie-Debut}

\clearpage
\section*{Tables}


\begin{table}[!ht]
\begin{center}
\begin{tabular}{| c | c | c |}
\hline
Model type	& Resistive										& Diffusive  \\
\hline
No dendrite 	& \b{$R_m$}, \b{$C_m$}, \b{$R_e$}						& \b{$R_m$}, \b{$C_m$}, \b{$A$}, \b{$B$}, \b{$f_W$} \\
\hline
Dendrite 		& \b{$R_s$}, \b{$C_s$}, \b{$R_e$}, \b{$l_d$} 		& \b{$R_s$}, \b{$C_s$}, \b{$A$}, \b{$B$}, \b{$f_W$}, \b{$l_d$}\\
\hline
\end{tabular} 

\caption{Parameters used for the four models. \b{$l_d$} is the length
  of the dendritic compartment. \b{$R_e$} is the extracellular
  resistance.}

\label{parameters}
\end{center}
\end{table}


\begin{table}[!ht]
\begin{center}
\begin{tabular}{| c | c | c | c | c |}
	\hline
	Model type	& \b{$Z_i = z_i l_s$} 	& \b{$Z_e$} & \b{$z_e^{(m)}$}	& \b{$\kappa_\lambda^2$}  \\
	\hline
	Standard 		& \b{$r_i l_s$} 	& \b{negligible} & 	\b{negligible}						& \b{$\frac{r_i}{r_m}(1+i\omega\tau_m)$}\\
	\hline
	Diffusive			& \b{$\frac{r_i}{(1+i)\sqrt{\omega}}l_s$} 	& \b{$\frac{A+iB}{1 + \sqrt{i\omega / \omega_W}}$} & \b{$\frac{0.5\tau_m}{2\pi a C_m (1+i)\sqrt{\omega}}$}	& \b{$\frac{z_i(1+i\omega\tau_m) }{r_m[1+\frac{z_e^{(m)}}{r_m}(1+i\omega\tau_m]}$} \\
	\hline
\end{tabular} 

\caption{Intermediate variables.  \b{$Z_i$} and \b{$Z_e$} are
  respectively the intracellular and extracellular impedances,
  \b{$z_e^{(m)}$} is the input resistance of the extracellular medium
  seen by the dendrite (in $\Omega/m$), \b{$\kappa_\lambda$} is the
  cable parameter of the system.  Constants: \b{$r_i$} and \b{$r_e$}
  are the linear density of resistance in the cytoplasm and
  extracellular medium (estimated to $28\times 10^9$ and $18\times
  10^9$ $\Omega/m$), \b{$z_i$} is the linear density of impedance in
  the dendrite, \b{$l_s$} and \b{$l_d$} are the length of the soma and
  dendrite, \b{$a$} is the diameter of the dendrite, \b{$\tau_m$} is
  the time constant of the membrane ($\sim$10~ms), \b{$C_m$} is the
  specific membrane capacitance ($10^{-2}~F/m^2$). We also have
  \b{$r_m = \tau_m/2\pi a _mC_m$}, where \b{$1/r_m$} is the linear
  density of membrane conductance (S/m).}

\label{formulas}
\end{center}
\end{table}


\clearpage


\section*{Figures}

\begin{figure}[!ht]


  \begin{center}
   \includegraphics[width=0.8\columnwidth]{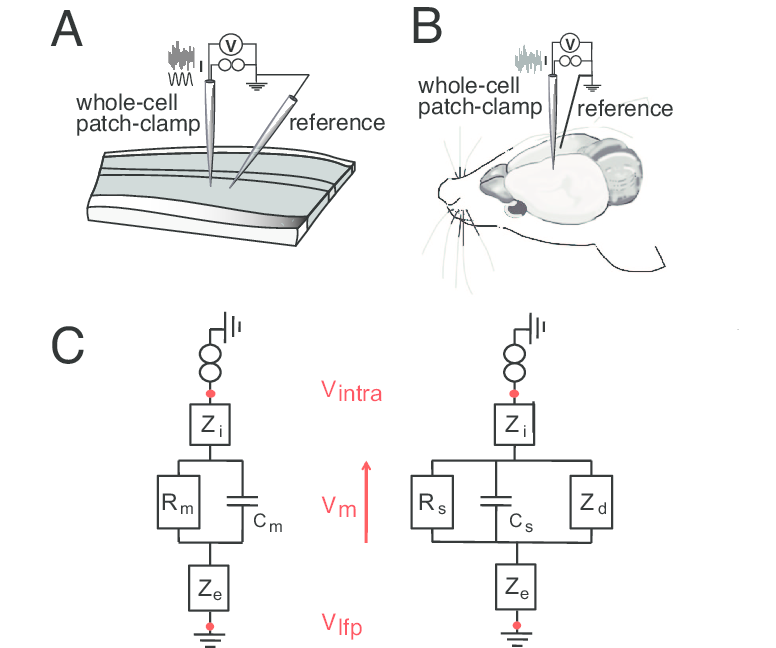}
  \end{center}

  \caption{\label{principle} Scheme of the experimental setup for
    measuring the global intracellular impedance.  {\bf A-B}:
    Placement of electrodes {\it in vitro} and {\it in vivo}.  In each
    case, a cell is recorded in patch-clamp whole-cell configuration
    using a micropipette, where the reference electrode is a
    micropipette located in the extracellular medium at a short
    distance from the recording (A), or a silver wire in the
    contralateral hemisphere (B).  {\bf C}: Equivalent circuits
    corresponding to this experimental setup, in two different
    configurations: the cell is either considered as a single
    compartment (left), or with dendrites (right), resulting in a
    slightly more complicated circuit.  The membrane is modeled as a
    RC-circuit, where \b{$R_m$} and \b{$C_m$} are the global membrane
    resistance and capacitance of the cell (left), respectively, or
    \b{$R_s$} and \b{$C_s$} are the resistance and capacitance of the
    somatic membrane (right).  \b{$Z_i$} is the macroscopic
    intracellular impedance (including the electrode-cytosol
    interface), and \b{$Z_e$} is the macroscopic extracellular
    impedance. In the resistive models, \b{$Z_e$} is modeled as a
    simple resistance \b{$R_e$}; in the diffusive models, it is a
    function of three parameters (\b{$A$}, \b{$B$} and \b{$k$},
    respectively scaling the modulus, phase and cutoff frequency of
    the diffusive impedance; see Materials and Methods).  In the right
    circuit, \b{$Z_d$} is the equivalent input impedance of the
    dendritic tree seen by the soma; it is a function of $l_d$, the
    equivalent length of the dendritic tree.  In both circuits,
    \b{$V_{intra}$} represents the intracellular potential, and
    \b{$V_{lfp}$} the potential in extracellular space.}

\end{figure}


\clearpage


\label{lastpage}

\clearpage

\appendix

\section*{Supplementary material}

\setcounter{equation}{0}
    \numberwithin{equation}{section}

{\section{Appendix 1: Integrating the global intracellular
        impedance in models}}
\label{App1}

{In this appendix, we relate the impedance $z_e^{(m)}$ in the generalized
cable, to the impedance measurements reported in the present paper.  In the
generalized cable~\cite{BedDes2013}, the extracellular impedance was modeled 
by parameter $z_e^{(m)}$.  Starting from the expression of the extracellular 
impedance, \b{
\begin{equation}
       Z_{e} = \frac{z_e^{(m)}}{A_{soma}}=\frac{z_e^{(m)}}{4\pi R_{soma}} 
\end{equation}}
and condisering a single-compartment model, according to 
Eq.~\ref{Z_eq_nodend}, we can write \b{
\begin{equation}
     z_e^{(m)} \approx A_{soma}Z_e= A_{soma}[Z_{eq}(\omega)- \frac{R_m}{1+i\omega \tau_m}] ~ .
\label{eqb2}
\end{equation}}
Thus, by assuming a typical somatic membrane area, we can estimate
$Z_{e}$, and thus also estimate $z_e^{(m)}$. The other parameters,
$R_m$, $\tau_m$, $Z_{eq}$, can also be estimated from the present
measurements.}

\setcounter{equation}{0}
    \numberwithin{equation}{section}

{\section{Appendix 2: Establishing the linearity of the
        system}}
\label{App2}

{In this Appendix, we explain how to determine the linearity of
  the system, in temporal and frequency space.}

{\subsection{Linearity in Fourier frequency space}}

{In the Fourier frequency domain, the experiments show that the
  ratio \b{$\Delta V(\omega)~/~I^g(\omega)$} is a bounded
  function for variations around the resting membrane potential.  In
  these conditions, a sinusoid in current gives a sinusoid voltage
  with the same frequency, and with no additional peak in the
  spectrum.  This is true for relatively small variations (a few
  millivolts), keeping the membrane far away from spike threshold.  As
  shown in Fig.~\ref{principle}D, a combination of sine-wave currents
  generates a voltage power spectrum with peaks at the same position
  in frequency, and where no additional peak or harmonics appear.  We
  can say that in this case, the membrane potential of the neuron is
  linear in Fourier frequency space.  This implies that each component
  of this system in this space is also linear, and in particular, the
  V-I relation of ion channels in the membrane are linear, because the
  membrane capacitance is approximately constant (White, 1970).  This
  is an expression of Ohm's law, in which the ion channels are
  equivalent to a resistor, with no voltage-dependent effects (see
  Section~\ref{A2}).}

{To demonstrate this, we note that the ratio between \b{$V$} and
  \b{$I^g$} is a continuous bounded function in Fourier frequency
  space, with the constraint \b{$V(0)=g(\omega,0)=0$} for \b{$\omega
    \neq 0$} (the latter condition means that the neuron is at rest
  when the transmembrane current is zero).  In these conditions, we
  have \b{
\begin{equation}
V(\omega) = g(\omega, I^g(\omega)) ~ .
\end{equation}}
We can develop \b{$g$} in Taylor series relative to the current, because
the domain of definition of \b{$g$} is necessarily compact in experimental
situations.  Consequently, we can write: \b{
\begin{equation}
\begin{array}{ccl}
\Delta V(\omega) = V(\omega)-V(0) &=&  \frac{\partial g}{\partial I^g}(\omega,0)I^g(\omega) + \frac{1}{2!}\frac{\partial^2g}{\partial I^{g2}}(\omega,0)I^{g2}(\omega) + \cdots ~ \\\\\\
&=& b_1(\omega)I^g(\omega) + b_2(\omega)I^{g2}(\omega)+\cdots ~ .
\end{array}
\end{equation}}
The impedance is then given by:
\b{
\begin{equation}
Z(\omega)=\frac{\Delta V(\omega)}{I^g(\omega)}  =  
 b_1(\omega) + b_2(\omega)I^g(\omega)+\cdots ~ .
\end{equation}}
We can see that, if the spectrum \b{$I(\omega)$} is a discrete Fourier spectrum
composed of Dirac delta functions, then \b{$Z(\omega)$} cannot be a bounded
function when \b{$b_n \neq 0$} for \b{$n>1$}. Thus, we obtain \b{
\begin{equation}
\Delta V{\omega} = b_1(\omega) I^g(\omega)
\end{equation}}
when the ratio \b{$V(\omega) / I^g(\omega)$} is a bounded function.
In other words, the system is necessarily linear in Frequency space
because the V-I relation does not depend on the current amplitude.
Note that this independence is only true in the absence of
voltage-dependent conductances, so it can apply to the subthreshold
range, near the resting membrane potential.  Such a linear dependence
also implies that the position of spectral lines is necessarily
identical between \b{$V(\omega)$} and \b{$I^g(\omega)$}.}

{\subsection{Linearity of traditional V-I curves}
\label{A2}}

{We now address the question of whether the V-I relation of ion
  channels is linear when these channels are linear in Fourier
  frequency space, and {\it vice-versa}.}

{In general, for a membrane containing ion channels, we have: \b{
\begin{equation}
V = f(I) ~ , 
\end{equation}}
with \b{V = f(0)= cst} for zero current (resting membrane potential).}

{We can approximate $V$ as precise as we want using a polynomial
  of the current, because $V$ is necessarily a continuous function of 
this variable since the electric field is finite ($\vec{E}=-\nabla V$).  
This is by virtue of the Stone-Weierstrass theorem \cite{Rudin}, which
states that every continuous function defined over a closed and bounded
domain, can be approximated as close as we want by a polynomial.  
Thus, for a given population of ion channels, we can write  \b{
\begin{equation}
\Delta V(t) = V(t)-V(0)=  a_1I + a_2 I^2 + \cdots.
\end{equation}} 
If we express \b{$I$} as \b{$I(t)=e^{i\omega t}$}, we obtain
\b{
\begin{equation}
\Delta V(t) 
= a_1e^{i\omega t} + a_2e^{2i\omega t} +\cdots
\end{equation}}
such that the Fourier transform of the variations of $V$ around \b{$V(0)$}
generally gives a spectrum very different from that of the current.
Indeed, applying the Fourier transform gives
\b{
\begin{equation}
\Delta V(\omega)=a_1~\delta(\omega'-\omega) + a_2~\delta(\omega'-2\omega)+ \cdots 
\end{equation}}
Thus, it is necessary that \b{$\forall n >1$} we have \b{$a_n=0$} if 
we want that the position of the spectral lines of \b{$\Delta V(\omega)$} is 
the same as that of \b{$I(\omega)$}.}

{Moreover, it is evident that if the function \b{$f$} is linear,
  then the position of the spectral lines of \b{$\Delta V(\omega)$} is
  the same as that of \b{$I(\omega)$}.}

{Thus, the linearity in Fourier frequency space implies linearity
  of the V-I relation of the ion channels activated in the range of
  $V$ where $V=f(I)$. The linearity in Fourier frequency space
  constitutes a full condition of linearity, because the V-I relation
  can be more complex, for example $V=f(\omega,I)$.}


\begin{thebibliography}{99}

\bibitem{Herreras} Makarova, J., Gomez-Gala, M., and Herreras, O.
  2008.  Variations in tissue resistivity and in the extension of
  activated neuron domains shape the voltage signal during spreading
  depression in the CA1 in vivo. European Journal of Neuroscience {\it
    27}: 444-456.

\bibitem{Buzsaki} Buzs\'aki, G., Anastassiou, C., and Koch, C. 2012.
  The origin of extracellular fields and currents: EEG, ECoG, LFP and
  spikes. Nature Reviews Neurosci. {\it 13}, 407-420.

\bibitem{Ranck} Ranck, J. 1963. Analysis of specific impedance 
of rabbit cerebral cortex. Exp. Neurol. {\it 7}, 144-152.

\bibitem{Nicho2005} Nicholson, C. 2005. Factors governing diffusing
  molecular signals in brain extracellular space. J.  Neural Transm.
  {\it 112}, 29-44 .
  
\bibitem{Logothetis} Logothetis N.K., Kayser, C. and Oeltermann, A.
  2007. In vivo measurement of cortical impedance spectrum in
  monkeys : implications for signal propagation. Neuron {\it 55}:
  809-823.

\bibitem{Schwan1968} {Schwan, HP, 1968. Electrode polarization
    impedance and measurements in biological materials.  Ann. New York
    Acad. Sci. {\it 148}: 191-209.}

\bibitem{Geddes} {Geddes, LA, 1997. Historical evolution of
    circuit models for the electrode-electrolyte interface. Ann.
    biomed. engin. {\it 25}: 1-14.}

\bibitem{McAdams} {McAdams, ET and Jossinet, J., 2000. Nonlinear
    transient response of electrode-electrolyte interfaces. Med. Biol.
    Engin. Comp {\it 38}: 427-432.}

\bibitem{Schwan1966} {Schwan, HP, 1966. Alternating current
    electrode polarization. Biophysik {\it 3}: 181-201.}

\bibitem{Gabriel} Gabriel, S., Lau, R.W. and Gabriel, C. 1996. The
  dielectric properties of biological tissues : II. Measurements in
  the frequency range 10 Hz to 20 GHz. Phys. Med.  Biol. {\it 41},
  2251-2269.

\bibitem{Wagner} {Wagner, T., Eden, U. et al., 2014. Impact of
    brain tissue filtering on neurostimulation fields: A modeling
    study.  Neuroimage {\it 85}: 1048-1057.}
  
\bibitem{BedDes2010} B\'edard, C., Rodrigues, S., Roy, N., Contreras,
  D., and Destexhe A. 2010.  Evidence for frequency-dependent
  extracellular impedance from the transfer function between
  extracellular and intracellular potentials. J. Computational
  Neurosci. {\it 29}, 389-403.

\bibitem{Deg2010} Dehghani, N, B\'edard, C., Cash, S.S., Halgren, E.
  and Destexhe, A. 2010. Comparative power spectral analysis of
  simultaneous elecroencephalographic and magnetoencephalographic
  recordings in humans suggests non-resistive extracellular media.  J.
  Computational Neurosci. {\it 29}, 405-421.

\bibitem{Nelson2013} Nelson M., Bosch, C., Venance, L., and Pouget, P.
  2013. Microscale Inhomogeneity of Brain Tissue Distorts Electrical
  Signal Propagation. J. Neurosci. {\it 33}, 3502-3512.

\bibitem{Paille2013} Paille, V., Fino, E., Du, K., Morera-Herreras, T., Perez, S., Kotaleski, J. H., and Venance, L. GABAergic circuits control spike-timing-dependent plasticity (2013). J. Neurosci. {\it 33}, 9353-9363.
\bibitem{Nelson2008} Nelson, M.J., Pouget, P., Nilsen, E.A., Patten,
  C.D., \& Schall, J.D. 2008.  Review of signal distortion through
  metal microelectrode recording circuits and filters.  J. Neurosci.
  methods {\it 169}, 141-157.

\bibitem{Pods} Pods, J., Schoenke, J. and Bastian, P. 2013.
  Electrodiffusion models of neurons and extracellular space using the
  poisson-nernst-planck equations - numerical simulation of the intra
  and extracellular potential for an axon model. Biophys. J. {\it 105}
  242-254.

\bibitem{Warburg1} Warburg, E. 1899. Ueber das Verhalten sogenannter
  unpolarisir barer Elektroden gegen Wechselstrom. Wied. Ann.
  {\it 67}, 493-499.

\bibitem{Warburg2} Warburg, E. 1901. Ueber Die Polarisations kapazitaet
  des Platins. Ann. Phys. {\it 6}, 125-135.
  
\bibitem{Bisquert} Bisquert, J., Garcia-Belmonte, G.,
  Fabregat-Santiago, F. and Bueno, P. 1999. Theoretical models for
  AC impedance of finite diffusion layers exhibiting low frequency
  dispersion. J. Electroanalytical Chem., {\it 475} 152-163.
  
\bibitem{BedDes2009} B\'edard, C. and Destexhe, A. 2009. Macroscopic
  models of local field potentials and the apparent 1/f noise in brain
  activity. Biophys. J. {\it 96}, 2589-2603.
    
\bibitem{BedDes2011} B\'edard, C. and Destexhe, A. 2011. A
  generalized theory for current-source density analysis in brain
  tissue.  Physical Review E {\it 84}, 041909.

 
\bibitem{Pettersen} Pettersen, K.H. and Einevoll, G.T. 2008.
  Amplitude variability and extracellular low-pass filtering of
  neuronal spikes. Biophys J. {\it 94}, 784-802.

\bibitem{Linden} Lind\'en, H., Pettersen, K.H. and Einevoll, G.T.
  2010.  Intrinsic dendritic filtering gives low-pass power spectra
  of local field potentials. J. Comput. Neurosci. {\it 29}, 423-444.

\bibitem{Rall1962} Rall, W. 1962. Electrophysiology of a dendritic
  neuron model. Biophys J. {\it 2}, 145-167.

\bibitem{Rall1995} Rall, W. 1995. The theoretical foundations of
  dendritic function (Cambridge, MA: MIT Press).
      
\bibitem{Hines97} Hines ML and Carnevale NT. 1997. The NEURON
simulation environment. Neural Comput. {\it 9}, 1179-1209.

\bibitem{BedDes2013} B\'edard, C. and Destexhe, A. 2013. Generalized
  cable theory for neurons in complex and heterogeneous media.
    Physical Review E {\it 88}, 022709.

\bibitem{Robinson} Robinson, D. 1968. The electrical properties of
  metal microelectrodes. Proc. IEEE {\it 56}, 1065–1071.

\bibitem{Schwan} Schwan, H. 1968. Electrode polarization impedance
  and measurements in biological materials. Ann. New York Acad. Sci.
  {\it 148}, 191-209.

\bibitem{Nunez2005} Nunez, P.L. and Srinivasan, R. 2005. {\it
Electric Fields of the Brain. The Neurophysics of EEG} (2nd edition).
Oxford university press, Oxford, UK.

\bibitem{Hille2001} Hille, B. 2001. {\it Ionic Channels of Excitable
    Membranes}, Sinauer Sunderland, MA.

\bibitem{Mitzdorf} Mitzdorf, U. 1985. Current source-density method
  and application in cat cerebral cortex: investigation of evoked
  potentials and EEG phenomena.  Physiol. Reviews {\it 65},
  37-100.

\bibitem{Rudin} {Rudin, W. 1976. {\it Principles of mathematical
      analysis.}  McGraw-Hill, New York.}


\bibitem{White} {White, SH.1970.  A study of lipid bilayer
    membrane stability using precise measurements of specific
    capacitance.  Biophys J. {\it 10}, 1127-1148.  }




\label{Bibliographie-fin}
\end{thebibliography}
\end{document}